\documentclass[journal]{IEEEtran}

\usepackage{amsmath, graphicx, amssymb, epsfig, color, cite}
\usepackage{epstopdf}
\usepackage{array}
\usepackage{multirow}
\usepackage{enumerate}
\usepackage{booktabs}
\usepackage{verbatim}
\usepackage{optidef}
\usepackage{algorithm}
\usepackage{algpseudocode}
\usepackage{algcompatible}
\definecolor{blue}{rgb}{0,0,0}
\algblockdefx{FORP}{ENDFORP}[1]%
  {\textbf{for }#1 \textbf{do in parallel}}%
  {\textbf{end}}
\algblockdefx{FOR}{ENDFOR}[1]%
  {\textbf{for }#1 \textbf{do}}%
  {\textbf{end}}
\algblockdefx{IF}{ENDIF}[1]%
  {\textbf{if }#1 \textbf{then}}%
  {\textbf{end}}
\algblockdefx{FORALL}{ENDFORALL}[1]%
  {\textbf{for all}#1 \textbf{do}}%
  {\textbf{end}}
  
\algblockdefx{IF}{ENDIF}[1]%
  {\textbf{if }#1 \textbf{then}}%
  {\textbf{end}}
\algblockdefx{WHILE}{ENDWHILE}[1]%
  {\textbf{while }#1 \textbf{do}}
  {\textbf{end}}
  
\algnewcommand\algorithmicprocedure{\textbf{function}}
\algnewcommand\FUNC{\item[\algorithmicprocedure]}%
\algnewcommand\algorithmicendprocedure{\textbf{end function}}
\algnewcommand\ENDFUNC{\item[\algorithmicendprocedure]}%
\algnewcommand\algorithmicforeach{\textbf{for each}}
\algdef{S}[FOR]{FOREACH}[1]{\algorithmicforeach\ #1\ \algorithmicdo}

\usepackage{bm}
\usepackage{float}
\usepackage{makecell}
\usepackage{graphicx}
\usepackage{lipsum}
\usepackage{capt-of}

\makeatletter
\newcommand{\vast}{\bBigg@{4.5}}
\newcommand{\Vast}{\bBigg@{7.5}}
\makeatother

\begin{document}

    \title{UV3CP: Universal ViT-based 3D Channel Predictor for MIMO-OFDM Systems} 
    \author{Sungyoung Ha, Yujin Kang, Seonjung Kim, and Yo-Seb Jeon
    \thanks{This work was supported in part by the National Research Foundation of Korea (NRF) grant funded by the Korea government (MSIT) (No. RS-2024-00453301) and in part by Samsung Electronics Co., Ltd (IO251217-14722-01).} 
    \thanks{Sungyoung Ha, Yujin Kang, Seonjung Kim, and Yo-Seb Jeon are with the Department of Electrical Engineering, POSTECH, Pohang, Gyeongbuk 37673, South Korea (e-mail: sungyoungha@postech.ac.kr; yujinkang@postech.ac.kr; seonjung.kim@postech.ac.kr; yoseb.jeon@postech.ac.kr).}
    
    }
 
	\vspace{-2mm}
	
	\maketitle
	\vspace{-12mm}
 
    \begin{abstract}  
    Channel aging in time-division duplex multiple-input multiple-output orthogonal frequency division multiplexing (MIMO-OFDM) systems causes a mismatch between uplink-based channel state information and the downlink channel at the transmission instant. Although vision transformers (ViTs) are effective for channel prediction, their performance is highly sensitive to the patch size, whose optimum depends on the channel coherence region. Moreover, changing the patch size typically requires retraining because it alters the token grid and patch embedding operator. In this paper, we propose a universal ViT-based 3D channel prediction framework, called {\em UV3CP}, for MIMO-OFDM systems. The proposed UV3CP employs an efficient 3D ViT backbone to capture frequency, time, and spatial channel dependencies. First, based on second-order channel statistics, we derive a statistics-aware patch-size selection rule that selects axis-wise patch lengths using a mean-squared-error-based criterion, thereby matching the tokenization granularity to the dominant channel coherence while controlling token length and complexity. Second, we develop a statistics-aware patch-embedding resizing method that enables a single shared UV3CP to operate across different patch sizes without patch-size-specific retraining. Simulation results show that the proposed method improves robustness under various scenarios and patch-size mismatch while reducing the need to train and store multiple patch-size-specific ViT models.
    \end{abstract}

    \begin{IEEEkeywords}
    MIMO-OFDM, TDD, Channel aging, Channel prediction, Vision transformer, Embedding resizing
    \end{IEEEkeywords}

    \section{Introduction}\label{Sec:Intro}

    Multiple-input multiple-output orthogonal frequency division multiplexing (MIMO-OFDM) is a key physical-layer technology in modern broadband wireless systems, including 5G New Radio (NR), due to its ability to jointly exploit spatial multiplexing and frequency-selective transmission over wide bandwidths \cite{5G, 5Gto6G}. The performance of MIMO-OFDM systems, however, critically depends on the availability of accurate channel state information (CSI) for signal detection, beamforming, precoding, and resource allocation \cite{mimo_ofdm_ce, mimo_ofdm_ce_2}. In practice, CSI acquisition is degraded by user mobility, processing latency, and the limited density of reference signals on the time-frequency resource grid \cite{5g_standards}. As a result, the acquired CSI can become outdated before it is used, which leads to performance degradation in subsequent transmission and reception procedures.
    
    This issue is particularly important in time-division duplex (TDD) systems, where the base station commonly infers downlink (DL) CSI from uplink (UL) reference signals by exploiting channel reciprocity \cite{TDD_calibration, TDD, TDD_2}. When the channel varies rapidly, the actual DL channel at the transmission instant can differ significantly from the earlier UL-based channel estimate. This mismatch, commonly referred to as \emph{channel aging}, becomes more severe as the Doppler spread increases or the pilot-to-data latency grows \cite{channel_aging_1, channel_aging_2}. \textcolor{blue}{Therefore, channel prediction has attracted significant attention as an approach to alleviate channel aging by predicting future CSI from past channel observations over a given prediction horizon.}
    
    Classical channel prediction methods are largely model-based. Representative examples include autoregressive (AR) models, Kalman filtering, and linear minimum mean-square error (LMMSE) prediction, which exploit temporal or spatio-temporal channel correlation to extrapolate future CSI from past observations \cite{Autoregressive, kalman, mmse}. These methods are analytically tractable and often computationally efficient when accurate channel statistics or suitable parametric assumptions are available. \textcolor{blue}{However, multi-step prediction in MIMO-OFDM involves a high-dimensional channel history jointly varying across frequency, time, and antenna dimensions, for which restrictive model assumptions may limit prediction accuracy.}
    
    \textcolor{blue}{This motivates data-driven approaches that learn richer channel dynamics directly from data \cite{DL_cp, lstm_cp, transformer_cp, meta_cp, llm4cp, channelmamba, linformer, continuous_time_cp}. Recurrent neural networks, convolutional neural networks, and transformer-based architectures have all been investigated to learn channel dynamics directly from data without constraining the predictor to an explicit low-dimensional parametric model.} Among them, transformer-based methods are especially attractive because self-attention can capture long-range dependencies over structured channel observations \cite{transformer_cp}. In particular, vision transformer (ViT)-style architectures provide a natural representation for MIMO-OFDM channels by partitioning the time-frequency-space channel tensor into local 3D patches and modeling global interactions among the resulting patch embeddings \cite{vit, vit_ce}. This is well aligned with MIMO-OFDM channels, whose local 3D structure is strongly influenced by Doppler spread, delay spread, and array-domain correlation.

    Despite this promise, direct studies of ViT-style channel predictors for MIMO-OFDM systems remain limited. An important practical issue is that their performance can be highly sensitive to the 3D patch size. The patch size determines the locality represented by each token, the number of tokens processed by self-attention, and the overall complexity-performance tradeoff. More importantly, the most suitable patch size is not universal \cite{flexivit, supervision}. Since the effective coherence region of a wireless channel depends on mobility, delay spread, and propagation conditions, the patch size that performs well in one scenario can become suboptimal in another \cite{adafortitran, delay_doppler}. Therefore, a fixed-patch ViT can suffer from robustness degradation under scenario mismatch.
    
    A straightforward solution is to redesign or retrain the model for each patch size. However, this is inefficient in practice because changing the patch size alters both the token grid and the patch embedding operator, thereby requiring separate training and storage of multiple patch-size-specific models. The core difficulty is that the patch size should adapt to the channel coherence, whereas changing the patch size in a ViT typically prevents direct model reuse. This motivates the need for a universal framework that can both select an appropriate patch size according to channel statistics and adapt a single ViT model across different patch sizes without full retraining.
    
    Motivated by these challenges, we propose a universal ViT framework, called {\em UV3CP}, for 3D channel prediction in MIMO-OFDM systems. The most prominent feature of the proposed UV3CP is that it enables both scenario-adaptive patch selection and robust model reuse across patch-size changes. \textcolor{blue}{Importantly, UV3CP follows a hybrid design: the shared ViT learns the high-dimensional channel-prediction mapping directly from channel realizations, whereas second-order channel statistics are used to construct statistics-aware patch embeddings during multi-patch training and to guide patch selection and embedding adaptation at deployment. Therefore, the statistical model serves as an adaptation prior rather than as a structural constraint on the neural prediction mapping.} In UV3CP, we train a shared ViT backbone over multiple candidate patch sizes in the offline stage. Then, at deployment time, we select a statistics-aware target patch size and adapt the corresponding embedding components to the selected size. 
    To this end, we derive a statistics-aware patch-size selection rule that determines domain-wise patch lengths using a prediction-error-oriented criterion, thereby aligning the tokenization resolution with the dominant coherence structure of the channel while controlling the token sequence length and computational cost. 
    We also develop a statistics-aware patch-embedding resizing technique that enables a single base model to operate across multiple target patch sizes without retraining. These components enable both scenario-adaptive patch selection and robust model reuse across patch-size changes.
    
    The main contributions of this paper are summarized as follows.
    \begin{itemize}
        \item We introduce UV3CP, an efficient ViT-based 3D channel predictor that combines 3D patch tokenization, domain-separated attention, and patch-size sampling for robust MIMO-OFDM channel prediction.
        
        \item \textcolor{blue}{We develop a domain-wise patch-size design for ViT-based MIMO-OFDM channel prediction. The temporal and frequency patch lengths are adaptively selected using second-order channel statistics, while the spatial patch length is fixed and empirically evaluated for the adopted antenna-pair representation.}
    
        \item \textcolor{blue}{We develop a statistics-aware patch-embedding resizing method that enables a single UV3CP to operate across different patch sizes without patch-specific retraining.}

        \item We demonstrate through simulations that the proposed framework improves robustness to scenario and patch-size mismatch while reducing the need to train or store multiple patch-size-specific ViT models.
    \end{itemize}
    
    The remainder of this paper is organized as follows. Sec. \ref{Sec: system_model} introduces the system model and overview of the proposed UV3CP framework. Sec. \ref{Sec: ViT}, Sec. \ref{Sec: Patching}, and Sec. \ref{Sec: deployment} describe the basic architecture, statistics-aware patch adaptation strategy, and training/deployment procedures of the proposed framework, respectively. Simulation results are presented in Sec. \ref{Sec: simulation}, followed by conclusions in Sec. \ref{Sec: conclusion}.

    \vspace{2mm}
    {\em Notation:} Scalars, vectors, matrices, and tensors are denoted by lowercase letters (e.g., $a$), bold lowercase letters (e.g., $\mathbf{a}$), bold uppercase letters (e.g., $\mathbf{A}$), and calligraphic uppercase letters (e.g., $\mathcal{A}$), respectively. The sets of real and complex numbers are denoted by $\mathbb{R}$ and $\mathbb{C}$, respectively. $\Re{\cdot}$ and $\Im{\cdot}$ denote the real and imaginary parts of a complex-valued quantity, respectively. The superscripts $(\cdot)^{-1}$, $(\cdot)^*$, $(\cdot)^{\sf T}$, and $(\cdot)^{\sf H}$ denote the matrix inverse, complex conjugate, transpose, and Hermitian transpose, respectively. The operator $\mathbb{E}[\cdot]$ denotes expectation. For a scalar $a$, $|a|$ denotes its absolute value. For a vector $\mathbf{a}$, $||\mathbf{a}||_2$ denotes its Euclidean norm, and $[\mathbf{a}]_i$ denotes its $i$-th entry. For a matrix $\mathbf{A}$, $[\mathbf{A}]_{i,j}$ denotes its $(i,j)$-th entry, and $||\mathbf{A}||_{\mathrm{F}}$ denotes its Frobenius norm. For a tensor $\mathcal{A}$, $[\mathcal{A}]_{i_1,\ldots,i_N}$ denotes its $(i_1,\ldots,i_N)$-th entry. $\mathbf{I}_m$ denotes the $m\times m$ identity matrix. The Kronecker product is denoted by $\otimes$. $\mathcal{CN}(\boldsymbol{\mu},\mathbf{C})$ denotes the circularly symmetric complex Gaussian distribution with mean vector $\boldsymbol{\mu}$ and covariance matrix $\mathbf{C}$.

    \section{System Model and Overview}\label{Sec: system_model}
    
    \subsection{MIMO-OFDM Systems}\label{SubSec:MIMO}
    We consider a TDD MIMO-OFDM system in which a user equipment (UE) equipped with $N_t$ antennas transmits uplink (UL) signals to a base station (BS) equipped with $N_r$ antennas. The OFDM resource grid consists of $K$ subcarriers indexed by $k\in\{0,\ldots,K-1\}$ and $T$ OFDM symbols indexed by $n\in\{0,\ldots,T-1\}$. Assuming a sufficiently long cyclic prefix, the received signal at the BS on resource element (RE) $(n,k)$ is modeled as
    \begin{align}
    \mathbf{y}[n,k] = \mathbf{H}[n,k]\mathbf{x}[n,k] + \mathbf{v}[n,k],
    \label{eq: rx_model}
    \end{align}
    where $\mathbf{y}[n,k]\in\mathbb{C}^{N_r\times 1}$ is the received vector, $\mathbf{H}[n,k]\in\mathbb{C}^{N_r\times N_t}$ is the channel frequency response (CFR) matrix, $\mathbf{x}[n,k]\in\mathbb{C}^{N_t\times 1}$ is the transmitted vector, and $\mathbf{v}[n,k]\sim\mathcal{CN}(\mathbf{0},\sigma_v^2\mathbf{I}_{N_r})$ denotes additive white Gaussian noise (AWGN). We define the vectorized CFR as
    \begin{align}
    \mathbf{g}[n,k] \triangleq \mathrm{vec}(\mathbf{H}[n,k]) \in \mathbb{C}^{N_rN_t \times 1}.
    \label{eq: vec_g}
    \end{align}

    \subsection{Statistics of MIMO-OFDM Channels}\label{SubSec: Stats}
    In this work, we consider a standard statistical model to characterize the time-frequency and spatial correlations of the MIMO-OFDM channel.

    Let $h_{r,t}[n,k]\triangleq [\mathbf{H}[n,k]]_{r,t}$ denote the $(r,t)$-th element of $\mathbf{H}[n,k]$. We assume that the time-frequency correlation is common across antenna pairs, while spatial dependence is modeled separately through the spatial covariance introduced below. We define the time-frequency correlation as
    \begin{align}
    R[\Delta n,\Delta k] \triangleq
    \mathbb{E}\!\left[h_{r,t}[n,k]\;h_{r,t}^*[n+\Delta n,\,k+\Delta k]\right],
    \label{eq: tf_corr_def}
    \end{align}
    and assume wide-sense stationary uncorrelated scattering (WSSUS), so that $R[\Delta n,\Delta k]$ depends only on $(\Delta n,\Delta k)$. \textcolor{blue}{To obtain a tractable statistical representation for patch selection and embedding adaptation in Sec.~\ref{Sec: Patching}, we consider the separable approximation:
    \begin{align}
        R[\Delta n,\Delta k] \approx \sigma_h^2\rho_t[\Delta n]\rho_f[\Delta k],
        \label{eq:tf_separable}
    \end{align}
    where $\sigma_h^2=R[0,0]$, and $\rho_t[\cdot]$ and $\rho_f[\cdot]$ denote the normalized temporal and frequency correlation functions satisfying $\rho_t[0]=\rho_f[0]=1$. This separability is an additional modeling approximation beyond WSSUS and decomposes the time-frequency correlation into temporal and frequency components.} 
    Note that $\rho_t[\Delta n]$ and $\rho_f[\Delta k]$ in \eqref{eq:tf_separable} can be obtained from the classical Jakes model and the channel power delay profile (PDP), respectively \cite{tf_correlation}.

    \textcolor{blue}{To characterize spatial dependence, we define the spatial covariance of the vectorized MIMO channel as
    \begin{align}
        \mathbf{R}_s \triangleq \mathbb{E}\left[ \mathbf{g}[n,k]\mathbf{g}[n,k]^{\mathsf H} \right].
        \label{eq: spatial_cov}
    \end{align}
    The spatial correlation is determined by the power angle spectrum (PAS) and the corresponding antenna geometry \cite{spatial_correlation}. We retain the general spatial covariance $\mathbf{R}_s$ without imposing separability between the transmit and receive antenna dimensions.
    Using the general spatial covariance $\mathbf R_s$, we further approximate the covariance of the vectorized CFR across time and frequency offsets in a separable form as
    \begin{align}
        \mathbf{R}_g[\Delta n,\Delta k] &\triangleq \mathbb{E}\!\left[ \mathbf{g}[n,k] \mathbf{g}[n+\Delta n,k+\Delta k]^{\mathsf H}\right] \nonumber\\
        &\approx \rho_t[\Delta n]\rho_f[\Delta k]\mathbf{R}_s.
        \label{eq:channel_covariance}
    \end{align}
    The resulting representation provides a tractable description of the dominant correlation structure across the time, frequency, and spatial domains.}
    
    {\color{blue}In practical systems, these second-order statistics can be updated from recent pilot-based CFR observations over a local stationarity interval. For example, the temporal and frequency correlations can be constructed from estimated Doppler and PDP information, respectively, while the spatial covariance can be obtained from sample covariances of the CFR estimates.}

    \begin{figure}[t]
        \centering
        \includegraphics[width=8.8cm]{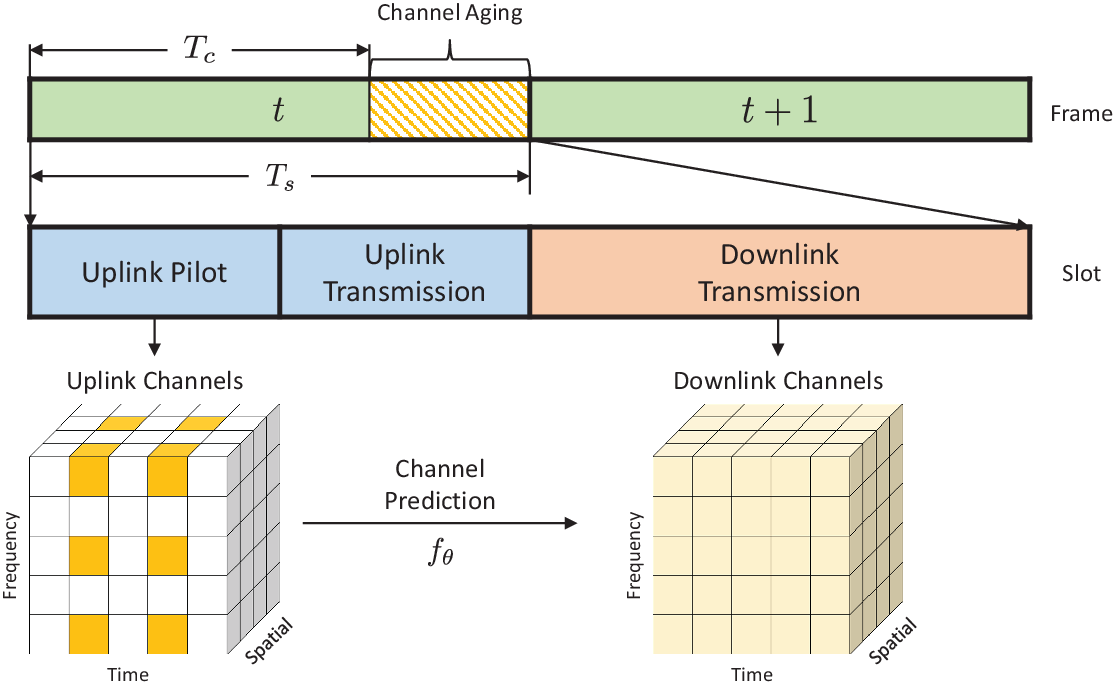}
        \caption{An illustration of channel aging in a TDD MIMO-OFDM frame, where future downlink channels are predicted from past uplink channel observations.}
        \vspace{-3mm}
        \label{fig:tdd_framework}
    \end{figure}

    \begin{figure*}[t]
        \centering
        \includegraphics[width=17cm]{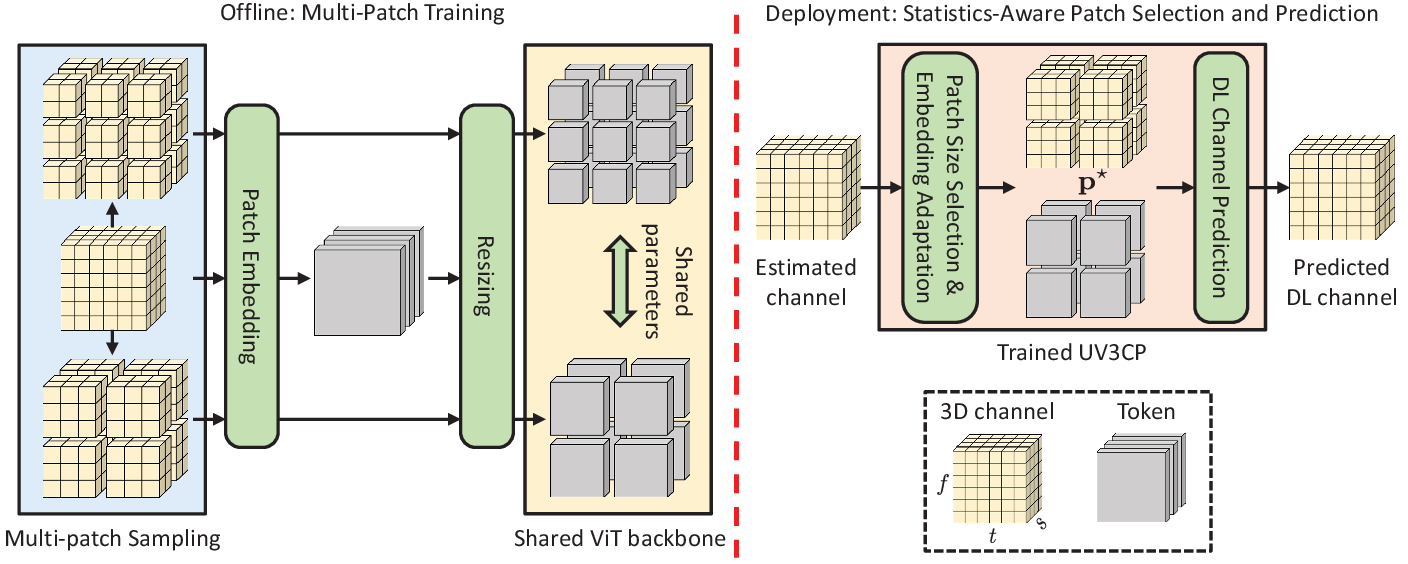}
        \caption{Overall framework of the proposed UV3CP. The offline stage trains a shared ViT backbone using multiple patch sizes, while the deployment stage selects a statistics-aware target patch size, adapts the embedding components, and predicts the future DL channel.}
        \vspace{-3mm}
        \label{fig: proposed framework}
    \end{figure*} 
    
    \subsection{Pilot-Based Channel Estimation}\label{SubSec:CE}
    We consider a TDD scenario, illustrated in Fig. \ref{fig:tdd_framework}, in which the time axis is organized into frames, each consisting of multiple slots with flexible UL and DL allocation at the symbol level \cite{5g_standards}. In this scenario, the UL channels are estimated using uplink pilot (reference) signals, and the DL channels are then inferred from the UL channel estimates based on UL-DL channel reciprocity. A typical UL channel estimation procedure in the TDD scenario is described below.


    Let $\mathcal{P}_{\mathrm{pilot}}$ denote the set of uplink pilot REs. For a pilot RE $(n,k)\in\mathcal{P}_{\mathrm{pilot}}$, we estimate $\mathbf{H}[n,k]$ by collecting $M$ pilot observations in a local neighborhood $\mathcal{N}_0=\{(n_i,k_i)\}_{i=1}^{M}\subseteq\mathcal{P}_{\mathrm{pilot}}$ around $(n,k)$. We stack the received and transmitted pilot symbols as
    \begin{align}
    \mathbf{Y}_p &\triangleq [\mathbf{y}[n_1,k_1],\ldots,\mathbf{y}[n_M,k_M]] \in \mathbb{C}^{N_r\times M}, \\
    \mathbf{X}_p &\triangleq [\mathbf{x}[n_1,k_1],\ldots,\mathbf{x}[n_M,k_M]] \in \mathbb{C}^{N_t\times M}.
    \label{eq:pilot_stack}
    \end{align}
    If the channel is approximately constant over the neighborhood, i.e., $\mathbf{H}[n_i,k_i]\approx \mathbf{H}[n,k]$ for all $(n_i,k_i)\in\mathcal{N}_0$, then the stacked received signal over the neighborhood can be approximated as
    \begin{equation}
    \mathbf{Y}_p \approx \mathbf{H}[n,k]\mathbf{X}_p + \mathbf{V}_p,
    \end{equation}
    where $\mathbf{V}_p\in\mathbb{C}^{N_r\times M}$ stacks the corresponding noise vectors. This corresponds to a locally pooled LS estimator, where neighboring pilot observations are jointly used under the local-constancy assumption. The least-squares (LS) estimate is
    \begin{equation}
    \hat{\mathbf{H}}[n,k]
    = \mathbf{Y}_p\mathbf{X}_p^{\mathsf H}\big(\mathbf{X}_p\mathbf{X}_p^{\mathsf H}\big)^{-1},
    \label{eq:ls_ce}
    \end{equation}
    provided that $\mathrm{rank}(\mathbf{X}_p)=N_t$, which is typically ensured by $M\ge N_t$ and appropriate pilot design. After obtaining LS channel estimates at pilot locations, we interpolate them over the entire resource grid to obtain $\hat{\mathbf{H}}[n,k]$ for all REs.

    \subsection{3D Channel Prediction Problem}\label{SubSec:PredictionFormulation}

    In the TDD scenario, the UL and DL channels are related through channel reciprocity enabled by calibration \cite{TDD_calibration, TDD, TDD_2}. However, as illustrated in Fig. \ref{fig:tdd_framework}, temporal channel variations can cause the DL channel at the transmission instant to deviate from earlier UL-based channel estimates, resulting in {\em channel aging}. This effect becomes more severe when the channel coherence time is short relative to the delay between UL channel estimation and DL transmission. To mitigate channel aging, we consider a 3D channel prediction problem that predicts $L$ future DL channels across the time, frequency, and spatial domains from $P$ past UL channel estimates.

    We first define the true CFR vector of the $n$-th OFDM symbol by stacking the CFRs over all subcarriers as
    \begin{align}
        \mathbf{h}_n \triangleq \big[\mathbf{g}[n,0]^{\mathsf T}, \ldots, \mathbf{g}[n,K-1]^{\mathsf T} \big]^{\mathsf T} \in \mathbb{C}^{d\times 1},
        \label{eq: h_n_def}
    \end{align}
    where $d=KN_rN_t$. Since only pilot-based channel estimates are available in practice, the corresponding observation vector is defined as
    \begin{align}
        \hat{\mathbf{h}}_n \triangleq \big[ \hat{\mathbf{g}}[n,0]^{\mathsf T}, \ldots, \hat{\mathbf{g}}[n,K-1]^{\mathsf T} \big]^{\mathsf T}\in \mathbb{C}^{d\times 1},
        \label{eq: h_hat_def}
    \end{align}
    where $\hat{\mathbf{g}}[n,k] \triangleq\mathrm{vec}(\hat{\mathbf{H}}[n,k]) \in \mathbb{C}^{N_rN_t\times 1}$.
    
    Let $f_{\theta}(\cdot)$ denote the 3D channel prediction model parameterized by $\theta$. Given the past $P$ observation vectors, the predictor estimates the future $L$ CFR vectors as
    \begin{align}
        \big( \tilde{\mathbf{h}}_{n+1}, \ldots, \tilde{\mathbf{h}}_{n+L}\big) = f_{\theta} \big(     \hat{\mathbf{h}}_{n-P+1}, \ldots,\hat{\mathbf{h}}_{n}\big),
        \label{eq: pred_map}
    \end{align}
    where $\tilde{\mathbf{h}}_{n+\ell}$ denotes the predicted CFR vector at the $(n+\ell)$-th OFDM symbol for $\ell=1,\ldots,L$.
    
    The prediction model is trained by minimizing the NMSE over the prediction horizon:
    \begin{align}
        \min_{\theta}\;\mathcal{L}(\theta) \triangleq\mathbb{E}\left[ \frac{\sum_{\ell=1}^{L} \left\|      \mathbf{h}_{n+\ell} - \tilde{\mathbf{h}}_{n+\ell}\right\|^{2}}{\sum_{\ell=1}^{L}\left\|    \mathbf{h}_{n+\ell}\right\|^{2}} \right].
        \label{eq: nmse_loss}
    \end{align}
    Since $\|\mathbf{h}_n\|^2 =\sum_{k=0}^{K-1}\|\mathbf{H}[n,k]\|_{\mathrm{F}}^2$, the objective in \eqref{eq: nmse_loss} is equivalently the normalized Frobenius MSE over all subcarriers and antenna pairs.

    \subsection{Overview of the Proposed UV3CP}

    ViT is well suited to MIMO-OFDM channel prediction because the channel history naturally forms a frequency-time-space tensor with structured correlations across the three domains. \textcolor{blue}{Treating every channel coefficient as an individual token would result in a long sequence and high attention cost. We therefore use 3D patches to aggregate channel coefficients along the frequency, time, and vectorized antenna-pair dimensions, reducing the sequence length while allowing self-attention to model interactions among channel regions.} However, the effectiveness of this representation depends on the patch size, since the channel coherence structure varies with mobility, delay spread, and spatial scattering. Consequently, a fixed patch size can become suboptimal under scenario mismatch.

    Motivated by this issue, we propose a ViT-based 3D channel predictor, called {\em UV3CP}, which combines a shared ViT backbone with statistics-aware patch adaptation. As illustrated in Fig. \ref{fig: proposed framework}, UV3CP is trained over multiple candidate patch sizes and, at deployment, selects a target patch size according to the current channel statistics and adapts the corresponding embedding components before predicting the future DL channel. The model architecture, patch adaptation strategy, and training/deployment procedures are described in Secs. \ref{Sec: ViT}, \ref{Sec: Patching}, and \ref{Sec: deployment}, respectively.
    
    \section{Basic Architecture of the Proposed UV3CP}\label{Sec: ViT}
    
    This section presents the basic architecture of UV3CP, which converts complex-valued channel observations into a two-channel real tensor and partitions it into 3D patches for token embedding. The resulting tokens are processed by a domain-separated transformer encoder along the frequency, time, and spatial domains, followed by a convolutional refinement head that produces the future channel tensor.

    \subsection{Input Representation and 3D Patch Embedding}
    \label{SubSec:Patching}
    
    Using the pilot-based channel estimate in Sec.~\ref{SubSec:CE}, we define the estimated vectorized CFR on each RE as
    \begin{align}
        \hat{\mathbf{g}}[n,k]
        \triangleq \mathrm{vec}(\hat{\mathbf{H}}[n,k])
        \in\mathbb{C}^{N_rN_t \times 1}.
    \end{align}
    Given the current OFDM symbol index $n$, we form a 3D tensor from the past $P$ estimated CFRs as
    \begin{align}
        [\hat{\mathcal{H}}_{n}]_{k,\tau,:}
        \triangleq \hat{\mathbf{g}}[n-P+\tau,k],
        \label{eq:Hhat_tensor}
    \end{align}
    where $k\in\{0,\ldots,K-1\}$ and $\tau\in\{1,\ldots,P\}$, so that
    $\hat{\mathcal{H}}_{n}\in\mathbb{C}^{K\times P\times N_rN_t}$.
    
    To process complex-valued inputs with real-valued neural networks, we use a two-channel representation. Define $\Phi:\mathbb{C}\to\mathbb{R}^{2}$ by
    $\Phi(z)\triangleq[\Re\{z\},\Im\{z\}]^{\mathsf T}$ and extend it elementwise to tensors. The resulting real-valued input tensor is
    \begin{align}
        \mathcal{X}_n
        \triangleq \Phi(\hat{\mathcal{H}}_n)
        \in\mathbb{R}^{K\times P\times N_rN_t\times 2}.
        \label{eq: Xn_real}
    \end{align}
    
    We partition $\mathcal{X}_n$ into non-overlapping 3D patches of size $(p_f,p_t,p_s)$ in the frequency, time, and spatial domains, respectively, where $p_f$, $p_t$, and $p_s$ divide $K$, $P$, and $N_rN_t$. Let
    $\mathcal{X}_{i,j,\ell}\in\mathbb{R}^{p_f\times p_t\times p_s\times 2}$
    denote the $(i,j,\ell)$-th patch, where
    $i\in\{0,\ldots,G_f-1\}$,
    $j\in\{0,\ldots,G_t-1\}$, and
    $\ell\in\{0,\ldots,G_s-1\}$, with
    $G_f=K/p_f$, $G_t=P/p_t$, and $G_s=N_rN_t/p_s$.
    
    Each patch is mapped to a $D$-dimensional token via a shared linear projection:
    \begin{align}
        \mathbf{t}_{i,j,\ell}
        \triangleq
        \mathbf{W}_e\,\mathrm{vec}(\mathcal{X}_{i,j,\ell})
        +\mathbf{b}_e
        \in \mathbb{R}^{D},
        \label{eq: patch_embed}
    \end{align}
    where $\mathbf{W}_e\in\mathbb{R}^{D\times d_p}$,
    $\mathbf{b}_e\in\mathbb{R}^{D}$, and
    $d_p\triangleq 2p_fp_tp_s$.
    This projection can be implemented efficiently by a 3D convolution with kernel size and stride $(p_f,p_t,p_s)$ and $D$ output channels.
    
    The 3D token grid is flattened into a 1D sequence and combined with positional embeddings as
    \begin{align}
        \mathbf{z}_m^{(0)}
        \triangleq
        \mathbf{t}_m+\mathbf{p}_m,
        \quad
        m\in\{1,\ldots,N_{\mathrm{tok}}\},
        \label{eq: token_seq}
    \end{align}
    where $N_{\mathrm{tok}}=G_fG_tG_s$ and
    $\mathbf{p}_m\in\mathbb{R}^{D}$ denotes the positional embedding associated with index $m$.
    The index $m$ is determined by a fixed bijection between $(i,j,\ell)$ and $\{1,\ldots,N_{\mathrm{tok}}\}$.
    The resulting transformer input sequence is
    \begin{align}
        \mathbf{Z}^{(0)}
        \triangleq
        \big[
            (\mathbf{z}_1^{(0)})^{\mathsf T},
            \ldots,
            (\mathbf{z}_{N_{\mathrm{tok}}}^{(0)})^{\mathsf T}
        \big]^{\mathsf T}
        \in\mathbb{R}^{N_{\mathrm{tok}}\times D}.
        \label{eq: Z0_def}
    \end{align}

    \subsection{Efficient Transformer Encoder and Convolutional Refinement Head}

    To exploit the anisotropic correlation structure of MIMO-OFDM channels, we adopt domain-separated attention along the frequency, time, and spatial domains. These correlations are governed by different physical factors, namely delay spread, Doppler spread, and spatial scattering, respectively, as reflected in \eqref{eq:channel_covariance}. Compared with full self-attention over the entire 3D token grid, this factorized structure captures domain-wise dependencies more efficiently. The input token matrix $\mathbf{Z}^{(0)}\in\mathbb{R}^{N_{\mathrm{tok}}\times D}$ is processed by a stack of $N_{\mathrm{enc}}$ transformer encoder layers. For the $r$-th layer, $\mathbf{Z}^{(r-1)}$ is first reshaped into a 4D token tensor as \begin{align} \mathcal{T}^{(r-1)} \triangleq \mathrm{reshape}\!\left(\mathbf{Z}^{(r-1)}\right) \in \mathbb{R}^{G_f\times G_t\times G_s\times D}. \end{align} The domain-separated attention is then sequentially applied along the three axes as $\mathcal{T}^{(r,f)}=\mathrm{DAttn}_f(\mathcal{T}^{(r-1)})$, $\mathcal{T}^{(r,t)}=\mathrm{DAttn}_t(\mathcal{T}^{(r,f)})$, and $\mathcal{T}^{(r,s)}=\mathrm{DAttn}_s(\mathcal{T}^{(r,t)})$, where $\mathrm{DAttn}_a(\cdot)$ denotes multi-head self-attention along axis $a\in\{f,t,s\}$ while keeping the other two token-grid indices fixed, together with residual connection and layer normalization. The resulting tensor is processed by a position-wise feed-forward block, yielding $\mathcal{T}^{(r)}=\mathrm{FFN}(\mathcal{T}^{(r,s)})$, and flattened as 
    \begin{align} 
        \mathbf{Z}^{(r)} = \mathrm{Flatten}\!\left(\mathcal{T}^{(r)}\right), \quad r\in\{1,\ldots,N_{\mathrm{enc}}\}. \label{eq: encoder_stack} 
    \end{align} 
    The final encoder output is $\mathbf{M}\triangleq\mathbf{Z}^{(N_{\mathrm{enc}})} \in\mathbb{R}^{N_{\mathrm{tok}}\times D}$. To obtain a fixed-dimensional latent representation, let $\mathbf{m}_m\in\mathbb{R}^{D}$ denote the transpose of the $m$-th row of $\mathbf{M}$. We then average the tokens as
    \begin{align} 
        \bar{\mathbf m} \triangleq \frac{1}{N_{\mathrm{tok}}} \sum_{m=1}^{N_{\mathrm{tok}}}\mathbf{m}_m \in\mathbb{R}^{D}. \label{eq: pooled_memory} 
    \end{align}
    This pooling provides a patch-size-independent interface to the prediction head. The pooled representation is linearly projected to a coarse real-valued feature tensor as
    \begin{align} 
        \mathcal{U}_n \triangleq \mathrm{reshape}\!\left( \mathbf{W}_p\bar{\mathbf m} +\mathbf{b}_p \right) \in \mathbb{R}^{K\times L\times N_rN_t\times C_h}, \label{eq: coarse_tensor} 
    \end{align} 
    where $\mathbf{W}_p\in\mathbb{R}^{KLN_rN_tC_h\times D}$ and $\mathbf{b}_p\in\mathbb{R}^{KLN_rN_tC_h}$ are trainable parameters, and $C_h$ denotes the number of hidden channels. The coarse tensor is refined by 3D convolutional layers as 
    \begin{align} 
        \tilde{\mathcal{X}}_{n+1:n+L} = \mathrm{Refine}\!\left(\mathcal{U}_n\right) \in \mathbb{R}^{K\times L\times N_rN_t\times 2}, \label{eq: refine_head} 
    \end{align} 
    where $\mathrm{Refine}(\cdot)$ maps the hidden features to the two-channel real representation of the predicted CFR tensor. The corresponding complex-valued CFR tensor is recovered as 
    \begin{align} 
        \tilde{\mathcal{H}}_{n+1:n+L} = \Phi^{-1}\!\left( \tilde{\mathcal{X}}_{n+1:n+L} \right) \in \mathbb{C}^{K\times L\times N_rN_t}. \label{eq: predicted_complex_tensor} 
    \end{align}
    Thus, the model predicts the next $L$ OFDM-symbol channels from the past $P$ observations in a multi-step manner.
    
    
    Since the appropriate patch size varies with the channel coherence structure, a fixed patch size may become suboptimal under changing channel conditions, motivating the statistics-aware patch adaptation in Sec.~\ref{Sec: Patching}.

    \section{Statistics-Aware Patch Adaptation \\of the Proposed UV3CP}\label{Sec: Patching}
    To achieve universal and adaptive operation of the proposed UV3CP architecture introduced in Sec.~\ref{Sec: ViT}, this section presents a statistics-aware patch adaptation strategy for UV3CP. The proposed strategy consists of patch-size selection (Sec.~\ref{SubSec: CorrAwarePatch}) and patch-embedding resizing (Sec. \ref{SubSec: uv_weight_resize}) techniques based on channel statistics.

    \subsection{Statistics-Aware Patch-Size Selection}\label{SubSec: CorrAwarePatch} 
    \textcolor{blue}{The patch size $(p_f,p_t,p_s)$ determines the local channel region represented by each token and the resulting number of tokens $N_{\rm tok}$, thereby affecting both prediction accuracy and computational complexity. From a classical signal-processing perspective, the frequency and temporal patch lengths can be viewed as local processing supports whose appropriate scales depend on the corresponding channel correlation or coherence regions. Motivated by the channel statistics in Sec. \ref{SubSec: Stats}, we therefore use $\rho_t[\cdot]$ and $\rho_f[\cdot]$ to guide the temporal and frequency tokenization granularities, respectively. Specifically, $p_t$ is determined using a prediction-oriented NMSE criterion, while $p_f$ is determined using a reconstruction-oriented spectral correlation criterion. Unlike $p_t$ and $p_f$, the spatial patch length $p_s$ is fixed in the considered implementation and evaluated empirically.}
    
    \subsubsection{Time-domain patch length}
    {\color{blue}
    The observation length $P$ of the proposed predictor is fixed and determines the total number of past OFDM symbols provided to the network. In contrast, the temporal patch length $p_t$ determines how these $P$ observations are partitioned into temporal tokens. To select $p_t$ according to the temporal channel statistics, we use the model-based LMMSE multi-step prediction error as a proxy for estimating a prediction-relevant temporal scale.

    The basic motivation is that, as a larger span of past channel observations is considered, temporally dependent observations generally provide additional information for predicting the future channel. Once extending the considered past span yields only a marginal reduction in the LMMSE prediction error, the additional older observations provide little incremental predictive information. We therefore use the corresponding temporal span as a statistical reference scale for temporal tokenization. Importantly, this scale is not interpreted as the observation length or the patch length itself, but as an auxiliary measure of the temporal extent over which past channel observations provide meaningful predictive information.
    
    To distinguish this auxiliary analysis from the fixed network observation length $P$, we introduce $Q\leq P$ as the number of most recent past observations considered in the LMMSE analysis. Recall $\mathbf{h}_n\in\mathbb{C}^{d\times 1}$ with $d=KN_rN_t$ in \eqref{eq: h_n_def}. For a given auxiliary past support $Q$ and prediction horizon $L$, define the stacked past and future vectors as
    \begin{align}
        \mathbf{z}_n^{(Q)} &\triangleq \big[\mathbf{h}_{n-Q+1}^{\mathsf T},\ldots,\mathbf{h}_{n}^{\mathsf T}\big]^{\mathsf T}\in\mathbb{C}^{dQ\times 1},
        \label{eq: z_q_def}\\
        \mathbf{z}_n^{(L)} &\triangleq \big[\mathbf{h}_{n+1}^{\mathsf T},\ldots,\mathbf{h}_{n+L}^{\mathsf T}\big]^{\mathsf T}\in\mathbb{C}^{dL\times 1}.
        \label{eq: z_l_def}
    \end{align}
    
    Motivated by the WSSUS assumption along with the separable second-order model in \eqref{eq:channel_covariance}, we assume that the vector process $\{\mathbf{h}_n\}$ shares the same normalized temporal correlation $\rho_t[\cdot]$ regardless of $n$:
    \begin{align}
        \mathbb{E}\left[\mathbf{h}_n\mathbf{h}_{n-\Delta}^{\sf H}\right] &= \rho_t[\Delta]\mathbf{R}_0,\quad \Delta\geq 0,
        \label{eq: hn_temporal_corr}
    \end{align}
    where $\mathbf{R}_0\triangleq\mathbb{E}[\mathbf{h}_n\mathbf{h}_n^{\sf H}]\in\mathbb{C}^{d\times d}$ captures the spatial-frequency covariance at a fixed OFDM symbol.
    
    Under \eqref{eq: hn_temporal_corr}, the covariance matrices of the stacked past and future vectors inherit a Kronecker structure between the temporal correlation matrices and $\mathbf{R}_0$. Accordingly, the covariance and cross-covariance matrices are given by
    \begin{align}
        \mathbf{R}_Q &\triangleq \mathbb{E}\left[\mathbf{z}_n^{(Q)}(\mathbf{z}_n^{(Q)})^{\sf H}\right] = \mathbf{T}_Q\otimes\mathbf{R}_0,
        \label{eq: stacked_covariances_1}\\
        \mathbf{R}_L &\triangleq \mathbb{E}\left[\mathbf{z}_n^{(L)}(\mathbf{z}_n^{(L)})^{\sf H}\right] = \mathbf{T}_L\otimes\mathbf{R}_0,
        \label{eq: stacked_covariances_2}\\
        \mathbf{R}_{LQ} &\triangleq \mathbb{E}\left[\mathbf{z}_n^{(L)}(\mathbf{z}_n^{(Q)})^{\sf H}\right] = \mathbf{S}_{LQ}\otimes\mathbf{R}_0,
        \label{eq: stacked_covariances_3}
    \end{align}
    where $\mathbf{T}_Q\in\mathbb{C}^{Q\times Q}$ and $\mathbf{T}_L\in\mathbb{C}^{L\times L}$ are Toeplitz matrices defined by $[\mathbf{T}_Q]_{i,j}=\rho_t[|i-j|]$ and $[\mathbf{T}_L]_{i,j}=\rho_t[|i-j|]$, respectively, and $\mathbf{S}_{LQ}\in\mathbb{C}^{L\times Q}$ has entries $[\mathbf{S}_{LQ}]_{\ell,q}=\rho_t[\ell+Q-q]$ for $\ell\in\{1,\ldots,L\}$ and $q\in\{1,\ldots,Q\}$.
    
    Let $\hat{\mathbf{z}}_n^{(L)}$ be the LMMSE estimate of $\mathbf{z}_n^{(L)}$ from $\mathbf{z}_n^{(Q)}$. Using \eqref{eq: stacked_covariances_1}--\eqref{eq: stacked_covariances_3}, the prediction error covariance is given by
    \begin{align}
        \mathbf{K}_{Q}^{(L)} &\triangleq \mathbb{E}\Big[(\mathbf{z}_n^{(L)}-\hat{\mathbf{z}}_n^{(L)})(\mathbf{z}_n^{(L)}-\hat{\mathbf{z}}_n^{(L)})^{\mathsf H}\Big] \nonumber\\
        &= \mathbf{R}_L-\mathbf{R}_{LQ}\mathbf{R}_Q^{-1}\mathbf{R}_{LQ}^{\mathsf H} \nonumber\\
        &= \big(\mathbf{T}_L-\mathbf{S}_{LQ}\mathbf{T}_Q^{-1}\mathbf{S}_{LQ}^{\mathsf H}\big)\otimes\mathbf{R}_0.
        \label{eq: KQL_final}
    \end{align}
    The corresponding average per-entry MSE is
    \begin{align}
        \mathrm{MSE}_{Q}^{(L)} &\triangleq \frac{1}{Ld}\mathrm{tr}\big(\mathbf{K}_{Q}^{(L)}\big) = \frac{\sigma_H^2}{L}\mathrm{tr}\big(\mathbf{T}_L-\mathbf{S}_{LQ}\mathbf{T}_Q^{-1}\mathbf{S}_{LQ}^{\mathsf H}\big),
        \label{eq: MSE_QL}
    \end{align}
    where $\sigma_H^2\triangleq\frac{1}{d}\mathrm{tr}(\mathbf{R}_0)$. The corresponding model-based normalized prediction error is $\mathrm{NMSE}_{Q}^{(L)} \triangleq \mathrm{MSE}_{Q}^{(L)}/\sigma_H^2$.
    
    For each prediction horizon $L$, we quantify the incremental prediction gain obtained by extending the considered past support from $Q-1$ to $Q$ as
    \begin{align}
        \Delta_t^{(L)}(Q) &\triangleq \frac{\mathrm{NMSE}_{Q-1}^{(L)}-\mathrm{NMSE}_{Q}^{(L)}}{\mathrm{NMSE}_{Q-1}^{(L)}}.
        \label{eq: temporal_incremental_gain}
    \end{align}
    The prediction-relevant temporal scale for horizon $L$ is then defined as
    \begin{align}
        Q^{\star}(L) &\triangleq \min\left\{Q-1\,\middle|\,\Delta_t^{(L)}(Q)\leq\eta_t\right\},
        \label{eq: Q_star_L_def}
    \end{align}
    where $Q\in\{2,\ldots,P\}$ and $\eta_t\in(0,1)$ denotes the marginal prediction-gain threshold. If no such $Q$ exists, we set $Q^{\star}(L)=P$.
    To obtain a temporal scale that is robust over the entire prediction horizon, we define
    \begin{align}
        Q_t^{\star} &\triangleq \max_{L\in\{1,\ldots,L_{t,\max}\}}Q^{\star}(L).
        \label{eq: Qt_star_robust}
    \end{align}
    The resulting $Q_t^{\star}$ characterizes the temporal span beyond which additional past channel observations provide only marginal incremental information for future-channel prediction. It is therefore interpreted as a prediction-relevant temporal scale rather than as the temporal patch length itself.
    
    The temporal patch length $p_t$ determines the number of consecutive OFDM symbols represented within each temporal token. We therefore use $Q_t^{\star}$ as a channel-statistics-based reference scale for selecting this tokenization granularity. Given the candidate temporal patch set $\mathcal{P}_t$, the implementable temporal patch length is selected as
    \begin{align}
        p_t^{\star} &\in \arg\min_{p\in\mathcal{P}_t,\,p\mid P}\left|p-Q_t^{\star}\right|,
        \label{eq: pt_from_Q}
    \end{align}
    where $p| P$ ensures non-overlapping tokenization of the fixed $P$-symbol observation window.
    
    We emphasize that the LMMSE criterion in \eqref{eq: temporal_incremental_gain}-\eqref{eq: Qt_star_robust} does not analytically optimize the ViT temporal patch length. Rather, it estimates a channel-dependent prediction-relevant temporal scale that serves as a physically motivated proxy for selecting $p_t$. Thus, $P$, $Q$, and $p_t$ have distinct roles: $P$ specifies the fixed amount of observed channel history, $Q$ is an auxiliary variable used to estimate the prediction-relevant temporal scale, and $p_t$ specifies the temporal tokenization granularity within the fixed observation window. The overall procedure is presented in Algorithm \ref{alg: Qt_star_robust}.
    
    \begin{algorithm}[t]
        \caption{Statistics-Aware Selection of Temporal Patch Length}
        \label{alg: Qt_star_robust}
        \begin{algorithmic}[1]
        \REQUIRE Temporal correlation $\rho_t[\cdot]$, observation length $P$, maximum prediction horizon $L_{t,\max}$, threshold $\eta_t$, candidate set $\mathcal{P}_t$
        \ENSURE Selected temporal patch length $p_t^{\star}$
        
        \FOR{$L=1$ to $L_{t,\max}$}
            \FOR{$Q=1$ to $P$}
                \STATE Construct $\mathbf{T}_Q$, $\mathbf{T}_L$, and $\mathbf{S}_{LQ}$ from $\rho_t[\cdot]$
                \STATE Compute $\mathrm{NMSE}_{Q}^{(L)}$
            \ENDFOR
        
            \STATE $Q^{\star}(L)\leftarrow P$
            \FOR{$Q=2$ to $P$}
                \STATE $\Delta_t^{(L)}(Q)\leftarrow\left(\mathrm{NMSE}_{Q-1}^{(L)}-\mathrm{NMSE}_{Q}^{(L)}\right)/\mathrm{NMSE}_{Q-1}^{(L)}$
                \IF{$\Delta_t^{(L)}(Q)\leq\eta_t$}
                    \STATE $Q^{\star}(L)\leftarrow Q-1$
                    \STATE \textbf{break}
                \ENDIF
            \ENDFOR
        \ENDFOR
        
        \STATE $Q_t^{\star}\leftarrow\max_{L\in\{1,\ldots,L_{t,\max}\}}Q^{\star}(L)$
        \STATE $p_t^{\star}\leftarrow\arg\min_{p\in\mathcal{P}_t,\,p\mid P}|p-Q_t^{\star}|$
        \STATE \textbf{return} $p_t^{\star}$
        \end{algorithmic}
    \end{algorithm}
    }
    \subsubsection{Frequency-domain patch length}
    While the temporal patch length above is determined using a prediction-oriented criterion, the frequency axis does not correspond to causal channel evolution. Instead, the frequency patch length should reflect the local spectral coherence across adjacent subcarriers. We therefore determine the frequency patch length using a local reconstruction criterion based on the frequency-domain correlation function. Specifically, we evaluate how accurately the channel at a reference subcarrier can be reconstructed from its neighboring subcarriers within a symmetric local frequency window.
    
    For a symmetric neighborhood with half-width $M$, define $\mathcal{S}_f(M)\triangleq\{-M,\ldots,-1,1,\ldots,M\}$, and let $\{s_i\}_{i=1}^{2M}$ be an arbitrary ordering of the elements in $\mathcal{S}_f(M)$. We consider reconstructing the channel at the reference subcarrier index $0$ from the neighboring subcarriers indexed by $\mathcal{S}_f(M)$. Under the same second-order model used for temporal patch selection, the covariance matrix of the neighboring observations is defined as
    \begin{equation}
    [\mathbf{R}_f(M)]_{i,j} \triangleq \rho_f[s_i-s_j], \quad i,j\in\{1,\ldots,2M\},
    \end{equation}
    and the cross-covariance vector between the reference subcarrier and the neighboring observations is defined as
    \begin{equation}
        [\mathbf{r}_f(M)]_i \triangleq \rho_f[-s_i], \quad i\in\{1,\ldots,2M\}.
    \end{equation}
    The normalized LMMSE reconstruction error for the reference subcarrier is then given by
    \begin{equation}
        \mathrm{NMSE}_f(M)\triangleq 1-\mathbf{r}_f^{H}(M)\mathbf{R}_f^{-1}(M)\mathbf{r}_f(M), ~ \mathrm{NMSE}_f(0)\triangleq1.
    \end{equation}
    The marginal gain obtained by increasing the local spectral span from $M-1$ to $M$ is defined as
    \begin{equation}
        \Delta_f(M)\triangleq \frac{\mathrm{NMSE}_f(M-1)-\mathrm{NMSE}_f(M)}{\mathrm{NMSE}_f(M-1)}.
    \end{equation}
    Accordingly, the statistically selected frequency half-width is chosen as
    \begin{equation}
        M_f^\star \triangleq \min\left\{M\in\{1,\ldots,M_{f,\max}\}\middle|\Delta_f(M)\le \eta_f\right\},
    \end{equation}
    where $\eta_f \in (0,1)$. If no such $M$ exists, we set $M_f^\star=M_{f,\max}$. Finally, since the corresponding local spectral support spans $2M_f^\star+1$ subcarriers including the reference subcarrier, the implementable frequency patch length is selected from the candidate set $\mathcal{P}_f$ as
    \begin{equation}
        p_f \in \arg\min_{p\in\mathcal{P}_f} \left|p-(2M_f^\star+1)\right|.
    \end{equation}
    
    \vspace{1mm}
    \subsubsection{Spatial-domain patch length} 
    {\color{blue}Unlike the temporal and frequency patch lengths, the spatial patch length is not adaptively selected from channel statistics. In the considered implementation, we fix $p_s=2$ as a fine-grained grouping of the vectorized antenna-pair coefficients. We note that adjacency in the resulting one-dimensional spatial axis does not necessarily imply physical antenna adjacency for an arbitrary array geometry. Therefore, $p_s=2$ is treated as an implementation choice rather than a universally optimal spatial configuration.}

    \subsection{Statistics-Aware Patch Embedding Resizing} \label{SubSec: uv_weight_resize}
    
    To enable a single shared ViT to operate across multiple patch sizes without patch-size-specific training, the patch-embedding operator must be adapted from a reference patch size to different target patch sizes. In the proposed UV3CP, a reference patch size $\mathbf{p}_0$ serves as an anchor for the shared embedding parameters, and the embedding associated with a target patch size $\mathbf{p}$ is instantiated from the reference embedding using second-order channel statistics. This statistics-aware resizing is employed during both multi-patch training and deployment. During training, it is applied to the patch size sampled at each iteration, whereas at deployment, it is applied to the target patch size selected according to the current channel statistics.
    
    Let $\mathbf{p}_0=(p_{f,0},p_{t,0},p_{s,0})$ denote the reference patch size and $\mathbf{p}=(p_f,p_t,p_s)$ denote a generic target patch size. Given the real-valued input tensor $\mathcal{X}_n\in\mathbb{R}^{K\times P\times N_rN_t\times 2}$, the shared ViT partitions the input tensor according to the selected patch size and maps each patch to a token through a linear patch-embedding operator. For notational convenience, the patch embedding associated with $\mathbf{p}$ is expressed as
    \begin{align}
        \mathbf{t}^{(\mathbf{p})} &= \mathbf{W}^{(\mathbf{p})}\mathbf{u}^{(\mathbf{p})}+\mathbf{b}^{(\mathbf{p})},
    \end{align}
    where $\mathbf{u}^{(\mathbf{p})}\in\mathbb{R}^{d_{\mathbf{p}}}$ is the flattened two-channel real patch vector and $\mathbf{W}^{(\mathbf{p})}\in\mathbb{R}^{D\times d_{\mathbf{p}}}$ is the corresponding embedding matrix. This formulation also includes Conv3D-based patch embedding, since convolution is linear and the flattening is introduced only for analysis and resizing.
    
    To adapt the reference model to a target patch size $\mathbf{p}$ without size-specific retraining, we construct a resized target embedding matrix from the reference embedding matrix. Let $\mathbf{u}_0\triangleq\mathbf{u}^{(\mathbf{p}_0)}\in\mathbb{R}^{d_{\mathbf{p}_0}}$ and $\mathbf{u}\triangleq\mathbf{u}^{(\mathbf{p})}\in\mathbb{R}^{d_{\mathbf{p}}}$ denote the reference and target patch vectors, respectively, and let $\mathbf{W}_0\triangleq\mathbf{W}^{(\mathbf{p}_0)}\in\mathbb{R}^{D\times d_{\mathbf{p}_0}}$ and $\mathbf{W}\triangleq\mathbf{W}^{(\mathbf{p})}\in\mathbb{R}^{D\times d_{\mathbf{p}}}$ denote the corresponding embedding matrices. Since the shared ViT operates on the two-channel real representation, both $\mathbf{u}_0$ and $\mathbf{u}$ are understood as real-valued vectorizations of the corresponding complex channel patches obtained by stacking their real and imaginary parts. Accordingly, all covariance and cross-covariance matrices in this subsection are defined in the resulting real vector space.
    
    We interpret $\mathbf{u}_0$ and $\mathbf{u}$ as two vectorized representations of the same underlying local channel content obtained at different tokenization resolutions induced by $\mathbf{p}_0$ and $\mathbf{p}$. Our objective is to construct $\mathbf{W}$ such that the target embedding preserves the token representation induced by the reference embedding as closely as possible in the mean-square sense. To this end, we introduce a linear target-to-reference operator that approximates the reference patch representation from the target patch representation as
    \begin{align}
        \mathbf{u}_0 &\approx \mathbf{B}\mathbf{u},
    \end{align}
    where $\mathbf{B}\in\mathbb{R}^{d_{\mathbf{p}_0}\times d_{\mathbf{p}}}$ denotes the target-to-reference transformation. For a given $\mathbf{B}$, the resized embedding matrix is determined by minimizing the mismatch between the reference embedding applied to the transformed target patch and the target embedding applied directly to the target patch:
    \begin{align}
        \mathbf{W}^{\star} &= \arg\min_{\mathbf{W}} \mathbb{E}\left[\left\|\mathbf{W}_0\mathbf{B}\mathbf{u}-\mathbf{W}\mathbf{u}\right\|^2\right].
    \end{align}
    Denoting the covariance of the target patch vector by $\mathbf{C}_u\triangleq\mathbb{E}[\mathbf{u}\mathbf{u}^{\sf T}]$, the above objective can be equivalently written as
    \begin{align}
        \mathbf{W}^{\star} &= \arg\min_{\mathbf{W}} \operatorname{tr}\left[(\mathbf{W}_0\mathbf{B}-\mathbf{W})\mathbf{C}_u(\mathbf{W}_0\mathbf{B}-\mathbf{W})^{\sf T}\right].
    \end{align}
    A minimizer of the above problem is given by
    \begin{align}
        \mathbf{W}^{\star} &= \mathbf{W}_0\mathbf{B}.
    \end{align}
    Hence, once a suitable target-to-reference operator is obtained, the resized embedding is given by the composition of $\mathbf{B}$ and the reference embedding matrix. The bias term is kept unchanged, i.e., $\mathbf{b}^{(\mathbf{p})}=\mathbf{b}^{(\mathbf{p}_0)}$, since the proposed adaptation resizes only the linear patch-embedding operator while preserving the learned token offset of the reference model.
    
    {\color{blue}
    We next construct the target-to-reference operator $\mathbf{B}$ by explicitly accounting for the second-order statistics of the channel patches. This provides a statistical counterpart to the practical pseudoinverse-based resizing rule used in flexible-patch ViTs such as FlexiViT \cite{flexivit}. In the proposed method, the resizing operator is determined from the statistical relationship between the reference and target channel patch representations.

    To characterize this relationship, we define the reference--target cross-covariance and the target-patch covariance as $\mathbf{C}_{0u}\triangleq\mathbb{E}[\mathbf{u}_0\mathbf{u}^{\sf T}]$ and $\mathbf{C}_{u}\triangleq\mathbb{E}[\mathbf{u}\mathbf{u}^{\sf T}]$, respectively. We choose $\mathbf{B}$ as the LMMSE linear estimator of the reference patch $\mathbf{u}_0$ from the target patch $\mathbf{u}$, yielding
    \begin{align}
        \mathbf{B}^{\star}=\arg\min_{\mathbf{B}}\mathbb{E}\!\left[\left\|\mathbf{u}_0-\mathbf{B}\mathbf{u}\right\|^2\right]=\mathbf{C}_{0u}\mathbf{C}_{u}^{\dagger},
    \end{align}
    where $(\cdot)^{\dagger}$ denotes the Moore--Penrose pseudoinverse. In general, $\mathbf{C}_{u}\succeq\mathbf{0}$. When $\mathbf{C}_{u}\succ\mathbf{0}$, the pseudoinverse reduces to the ordinary inverse, i.e., $\mathbf{C}_{u}^{\dagger}=\mathbf{C}_{u}^{-1}$; otherwise, it gives the corresponding minimum-norm LMMSE solution. Substituting $\mathbf{B}^{\star}$ into the embedding relation yields $\mathbf{W}^{\star}=\mathbf{W}_{0}\mathbf{C}_{0u}\mathbf{C}_{u}^{\dagger}$. Hence, the resized patch embedding is determined by the second-order relationship between the reference and target channel patches.
    
    We next specify how the required covariance and cross-covariance matrices are constructed when the reference and target patches have different supports. Let $\mathcal{Q}_0$ and $\mathcal{Q}$ denote the local coordinate sets corresponding to the reference patch $\mathbf{p}_0$ and target patch $\mathbf{p}$, respectively. Each coordinate specifies a relative frequency index, temporal index, and spatial antenna-pair index with respect to a common local patch origin. Thus, the two patches need not contain the same number of entries or admit a one-to-one correspondence between their discrete samples.
    
    Specifically, let $q_{0,i}=(\delta k_{0,i},\delta n_{0,i},s_{0,i})\in\mathcal{Q}_0$ denote the $i$-th coordinate of the reference patch and $q_j=(\delta k_j,\delta n_j,s_j)\in\mathcal{Q}$ denote the $j$-th coordinate of the target patch. Under the separable second-order approximation in Sec. II-B, the complex-domain reference--target cross-covariance is constructed as
    \begin{align}
        \left[\mathbf{C}_{0u}^{(\mathbb{C})}\right]_{i,j} = \rho_t[\delta n_j-\delta n_{0,i}]\rho_f[\delta k_j-\delta k_{0,i}][\mathbf{R}_s]_{s_{0,i},s_j}.
    \end{align}
    The entries of $\mathbf{C}_{00}^{(\mathbb{C})}$ and $\mathbf{C}_{u}^{(\mathbb{C})}$ are constructed analogously using pairs of coordinates from $\mathcal{Q}_0$ and $\mathcal{Q}$, respectively. Therefore, patches with different frequency, temporal, or spatial supports are aligned through their relative coordinates and second-order relationships rather than through direct one-to-one sample matching.
    
    For the spatial coordinate, we use a fixed vectorization of the MIMO channel matrix that preserves the correspondence of each spatial entry to a specific transmit-receive antenna pair. Hence, the spatial coordinates of the reference and target patches represent ordered antenna-pair coefficients rather than arbitrary unlabeled positions in a flattened vector.
    
    For further interpretation of the relation between the two patch supports, we define the reverse reference-to-target operator as the LMMSE linear estimator of $\mathbf{u}$ from $\mathbf{u}_0$:
    \begin{align}
        \mathbf{S}^{\star} = \arg\min_{\mathbf{S}} \mathbb{E}\!\left[\left\|\mathbf{u}- \mathbf{S}\mathbf{u}_0\right\|^2\right] = \mathbf{C}_{u0}\mathbf{C}_{00}^{\dagger},
    \end{align}
    where $\mathbf{C}_{00}\triangleq\mathbb{E}[\mathbf{u}_0\mathbf{u}_0^{\sf T}]$ and $\mathbf{C}_{u0}\triangleq\mathbb{E}[\mathbf{u}\mathbf{u}_0^{\sf T}]=\mathbf{C}_{0u}^{\sf T}$. Accordingly, the target patch can be represented as $\mathbf{u}=\mathbf{S}^{\star}\mathbf{u}_0+\mathbf{n}$, where $\mathbf{n}=\mathbf{u}-\mathbf{S}^{\star}\mathbf{u}_0$ is the residual component that cannot be linearly inferred from the reference patch. Note that $\mathbf{S}^{\star}$ and $\mathbf{B}^{\star}$ describe opposite mappings: $\mathbf{S}^{\star}$ maps the reference patch to the target patch, whereas $\mathbf{B}^{\star}$ maps the target patch to the reference patch. In general, the two operators are not inverses of each other.
    
    Because $\mathbf{S}^{\star}$ is the LMMSE linear estimator, the residual $\mathbf{n}$ is orthogonal to $\mathbf{u}_0$, i.e., $\mathbb{E}[\mathbf{n}\mathbf{u}_0^{\sf T}]=\mathbf{0}$. Its covariance is therefore
    \begin{align}
        \mathbf{C}_n = \mathbf{C}_{u} - \mathbf{S}^{\star}\mathbf{C}_{00} \left(\mathbf{S}^{\star}\right)^{\sf T}.
    \end{align}
    Consequently, $\mathbf{C}_{0u}=\mathbf{C}_{00}(\mathbf{S}^{\star})^{\sf T}$ and $\mathbf{C}_{u}=\mathbf{S}^{\star}\mathbf{C}_{00}(\mathbf{S}^{\star})^{\sf T}+\mathbf{C}_{n}$. Substitution into the target-to-reference LMMSE operator gives
    \begin{align}
        \mathbf{B}^{\star} = \mathbf{C}_{00} \left(\mathbf{S}^{\star}\right)^{\sf T} \left(\mathbf{S}^{\star}\mathbf{C}_{00} \left(\mathbf{S}^{\star}\right)^{\sf T} +\mathbf{C}_{n}\right)^{\dagger}.
    \end{align}
    The target embedding then follows directly from $\mathbf{W}^{\star}=\mathbf{W}_0\mathbf{B}^{\star}$.
    
    Under the separable second-order approximation, the reference-to-target relation can be constructed axis-wise. For $a\in\{f,t,s\}$, let $\mathbf{C}_{a,00}$ denote the covariance over the reference-patch coordinates in domain $a$, and let $\mathbf{C}_{a,u0}$ denote the corresponding target-reference cross-covariance. The domain-wise operator is $\mathbf{S}_{a}^{\star}=\mathbf{C}_{a,u0}\mathbf{C}_{a,00}^{\dagger}$. In the corresponding complex-domain representation, the multidimensional operator can then be written as $\mathbf{S}_{\mathbb C}^{\star}=\mathbf{S}_{f,\mathbb C}^{\star}\otimes\mathbf{S}_{t,\mathbb C}^{\star}\otimes\mathbf{S}_{s,\mathbb C}^{\star}$, with the ordering chosen consistently with the patch-vectorization convention. This factorization provides an efficient implementation of the covariance-based relation under the adopted separable approximation.
    
    The relationship between the proposed statistics-aware resizing and pseudoinverse-based resizing can be clarified through a special case. Consider an isotropic reference-patch distribution and a distortion-free reference-to-target relation, i.e., $\mathbf{C}_{00}=\sigma_u^2\mathbf{I}$ and $\mathbf{C}_{n}=\mathbf{0}$. Then,
    \begin{align}
        \mathbf{B}^{\star} = \sigma_u^2 \left(\mathbf{S}^{\star}\right)^{\sf T} \left( \sigma_u^2\mathbf{S}^{\star} \left(\mathbf{S}^{\star}\right)^{\sf T} \right)^{\dagger} = \left(\mathbf{S}^{\star}\right)^{\dagger}.
    \end{align}
    Accordingly, $\mathbf{W}^{\star}=\mathbf{W}_{0}(\mathbf{S}^{\star})^{\dagger}$. Under the matrix orientation adopted in this work, this corresponds to the pseudoinverse-based resize principle used in FlexiViT. Hence, this resizing can be interpreted as an isotropic and distortion-free special case of the proposed covariance-aware LMMSE formulation, whereas UV3CP retains the channel-dependent second-order relationship between the reference and target channel patches.

    Finally, the above formulation is implemented in the two-channel real representation used by the neural network. The required statistics are first constructed from the underlying complex-valued channel statistics. The complex-valued covariance matrices are converted to their real-valued representations to match the real-imaginary channel format used by the neural network.
    The same construction is applied to the cross-covariance matrices. When the real and imaginary components are stored in an interleaved order, fixed permutation matrices are additionally applied to match the actual patch-vectorization convention.
    The reference patch size $\mathbf{p}_0$ therefore serves as the anchor from which the embedding associated with any candidate target patch size is instantiated using the proposed statistics-aware resizing operator. This construction is used during both multi-patch training and deployment, while deployment additionally determines the target patch size according to the current channel statistics.}

    \section{Training and Deployment \\of the Proposed UV3CP}\label{Sec: deployment}
    {\color{blue}
    Building on the statistics-aware patch adaptation strategy in Sec. \ref{Sec: Patching}, UV3CP trains a single shared ViT backbone over multiple candidate patch sizes, with the embedding for each sampled patch size instantiated using the resizing method in Sec. \ref{SubSec: uv_weight_resize}. At deployment, the estimated channel statistics are used to select the target patch size and adapt the corresponding embedding without patch-size-specific retraining. Fig. \ref{fig: proposed framework} summarizes the overall offline training and online adaptation procedures.}

    \subsection{Training Stage: Multi-Patch Training with Statistics-Aware Embedding Adaptation}
    
    {\color{blue}Multi-patch training is a general strategy for supporting different tokenization granularities. UV3CP adopts this paradigm while using MIMO-OFDM channel second-order statistics for patch-embedding adaptation. Specifically, rather than learning an independent patch-embedding operator for every candidate patch size, UV3CP instantiates the embedding for each sampled patch size from a common reference embedding using the statistics-aware resizing method developed in Sec. \ref{SubSec: uv_weight_resize}.}
    
    We define a predefined candidate set of 3D patch sizes as
    \begin{align}
        \mathcal{P}_{\mathrm{patch}} = \left\{\mathbf{p}^{(1)},\mathbf{p}^{(2)},\ldots,\mathbf{p}^{(M)}\right\},
        \label{eq: patch_set}
    \end{align}
    where $\mathbf{p}^{(m)}=(p_f^{(m)},p_t^{(m)},p_s^{(m)})$ denotes the $m$-th candidate patch configuration in the frequency, time, and spatial domains.
    At each training iteration, one patch size $\mathbf{p}\in\mathcal{P}_{\mathrm{patch}}$ is sampled according to $\pi(\mathcal{P}_{\mathrm{patch}})$. The sampled patch size determines the tokenization grid, and the corresponding patch embedding is instantiated from the reference embedding using the statistics-aware resizing operator in Sec. \ref{SubSec: uv_weight_resize}. Thus, the shared ViT backbone is trained across multiple tokenization granularities and their corresponding statistics-aware embeddings.
    
    The shared parameters are learned by minimizing
    \begin{align}
    \boldsymbol{\theta}^{\star} = \arg\min_{\boldsymbol{\theta}} \mathbb{E}_{(\hat{\mathcal{H}},\mathcal{H}^{\mathrm{tar}}),\,\mathbf{p}\sim\pi(\mathcal{P}_{\mathrm{patch}})} \left[\mathcal{L}\left(f_{\boldsymbol{\theta}}(\hat{\mathcal{H}};\mathbf{p}),\mathcal{H}^{\mathrm{tar}}\right) \right], \label{eq: shared_training_obj}
    \end{align}
    where $f_{\boldsymbol{\theta}}(\hat{\mathcal{H}};\mathbf{p})$ denotes the proposed predictor associated with patch size $\mathbf{p}$, and $\mathcal{H}^{\mathrm{tar}}$ is the ground-truth future channel tensor.
    In this work, patch sizes are sampled uniformly from $\mathcal{P}_{\mathrm{patch}}$, while the transformer backbone and reference embedding parameters are shared across all candidates and the tokenization-dependent embedding is instantiated according to the sampled patch size and channel statistics. Consequently, UV3CP learns a single shared model that supports multiple patch resolutions.

    
    \subsection{Deployment Stage: Statistics-Aware Patch Adaptation}
    \begin{algorithm}[t]
        \caption{Training and deployment of the proposed UV3CP framework}
        \label{alg: universal_vit_training_deployment}
        \begin{algorithmic}[1]
        \REQUIRE Candidate patch set $\mathcal{P}_{\mathrm{patch}}$, reference patch size $\mathbf{p}_0$
        \ENSURE Trained shared model parameters $\boldsymbol{\theta}^{\star}$
        
        \STATE \textbf{Training:}
        \STATE Initialize the shared model and designate $\mathbf{p}_0$ as the reference patch size
        \FOR{each training batch $(\hat{\mathcal{H}},\mathcal{H}^{\mathrm{tar}})$}
            \STATE Sample $\mathbf{p}\sim \mathrm{Uniform}(\mathcal{P}_{\mathrm{patch}})$
            \STATE Construct the tokenization and embedding components for $\mathbf{p}$
            \STATE Compute $\tilde{\mathcal{H}}\leftarrow f_{\boldsymbol{\theta}}(\hat{\mathcal{H}};\mathbf{p})$
            \STATE Update $\boldsymbol{\theta}$ by backpropagation on $\mathcal{L}(\tilde{\mathcal{H}},\mathcal{H}^{\mathrm{tar}})$
        \ENDFOR
        
        \STATE \textbf{Deployment:}
        \STATE Collect recent channel observations and estimate channel statistics for patch selection
        \STATE Select $\mathbf{p}^{\star}$ using the correlation-aware patch selection rule
        \STATE Construct the reference-target cross-covariance and the target-patch covariance in the real-valued patch domain
        \STATE Form $\mathbf{B}^{\star}$ according to Sec. \ref{SubSec: uv_weight_resize}
        \STATE Instantiate $\mathbf{W}^{(\mathbf{p}^{\star})}$ from the reference embedding matrix
        \STATE Adapt the positional embedding to the token grid of $\mathbf{p}^{\star}$
        \STATE Compute $\tilde{\mathcal{H}}_{n+1:n+L}\leftarrow f_{\boldsymbol{\theta}^{\star}}(\hat{\mathcal{H}}_n;\mathbf{p}^{\star})$
        \end{algorithmic}
    \end{algorithm}
    {\color{blue}
    During deployment, the estimated channel statistics play two distinct roles in UV3CP: selecting an appropriate tokenization granularity for the current channel condition and adapting the patch embedding to the selected granularity. The BS first collects recent channel observations and estimates the second-order channel statistics required for patch adaptation. In particular, the temporal and frequency-domain correlation statistics are used for patch-size selection, while the patch-domain covariance statistics are used for embedding resizing.}
    Based on the estimated channel statistics, the BS determines the target 3D patch size
    \begin{align}
        \mathbf{p}^{\star} = \left(p_f^{\star},p_t^{\star},p_s^{\star}\right),
        \label{eq: target_patch_deploy}
    \end{align}
    using the statistics-aware patch-size selection strategy in Sec.~\ref{SubSec: CorrAwarePatch}. The selected patch size balances the channel coherence structure with the resulting token sequence length and transformer complexity.
    The BS then constructs the required second-order patch statistics and instantiates the patch embedding for $\mathbf{p}^{\star}$ from the shared reference embedding according to Sec.~\ref{SubSec: uv_weight_resize}. The positional embedding is adapted to the corresponding token grid.
    The input tensor $\mathcal{X}_n$ is tokenized using $\mathbf{p}^{\star}$ and processed by the instantiated embedding, the shared transformer backbone, and the prediction head to obtain the multi-step prediction $\tilde{\mathcal{H}}_{n+1:n+L}$.
    
    During training, the shared backbone is exposed to multiple candidate patch resolutions and their statistics-aware embeddings, whereas at deployment the current channel statistics determine the target patch configuration. Since only the tokenization-dependent components are adapted, UV3CP supports changing channel conditions without retraining or storing separate ViT models for different patch sizes. The overall algorithm is shown in Algorithm \ref{alg: universal_vit_training_deployment}. \textcolor{blue}{Using low-dimensional channel statistics for patch adaptation avoids repeated neural-network evaluation over all candidate patch configurations whenever the channel condition changes.}
    
    \begin{figure*}[t]
        \centering
        \begin{minipage}[t]{0.32\textwidth}
            \centering
            \includegraphics[width=\linewidth]
            {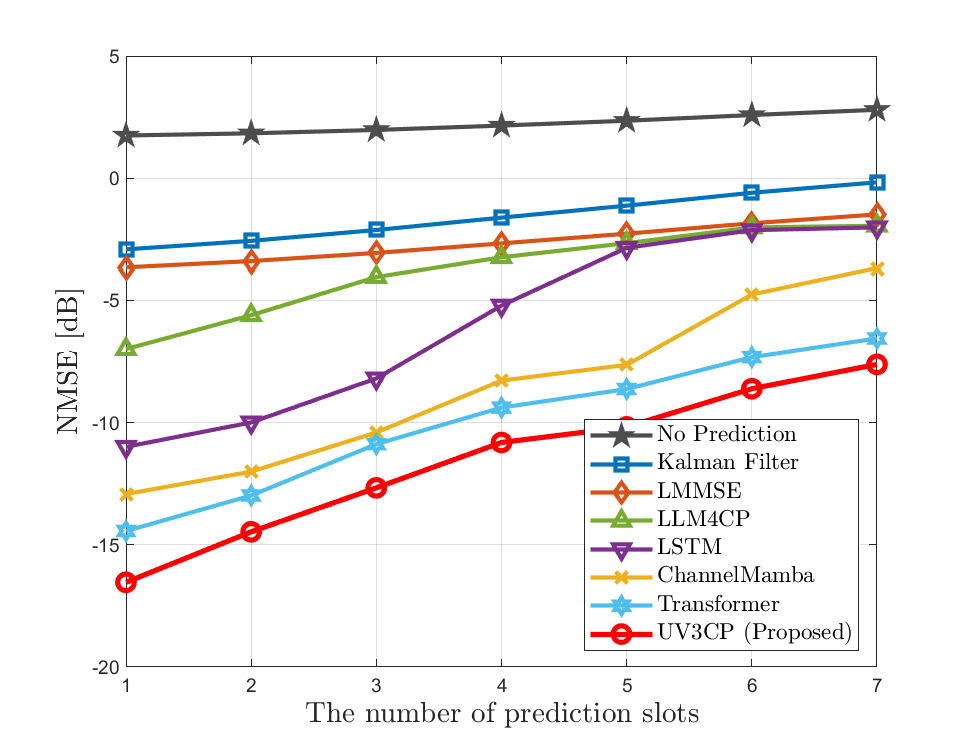}
            \caption{NMSE versus prediction horizon under a seen channel condition (CDL-A, $100$ ns delay spread, $5$ m/s velocity) at $E_b/N_0=-4\,\mathrm{dB}$.}
            \label{fig: NMSE_seen}
        \end{minipage}
        \hfill
        \begin{minipage}[t]{0.32\textwidth}
            \centering
            \includegraphics[width=\linewidth]
            {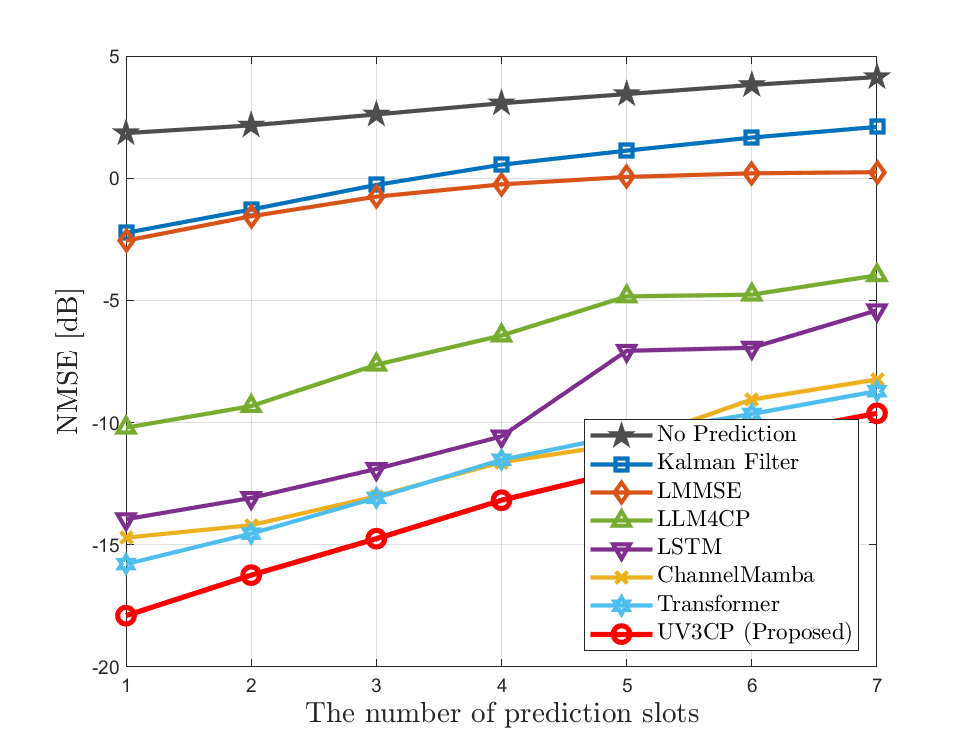}
            \caption{NMSE versus prediction horizon under an unseen channel condition (CDL-D, $200$ ns delay spread, $10$ m/s velocity) at $E_b/N_0=-4\,\mathrm{dB}$.}
            \label{fig: NMSE_unseen}
        \end{minipage}
        \hfill
        \begin{minipage}[t]{0.32\textwidth}
            \centering
            \includegraphics[width=\linewidth]
            {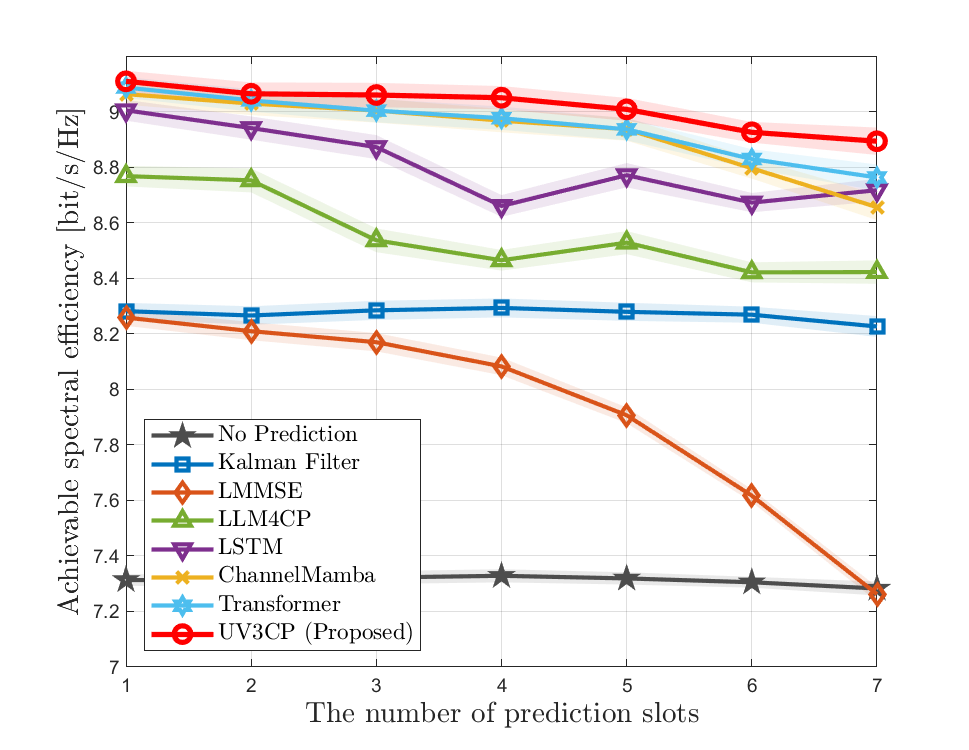}
            \caption{Achievable rate versus prediction horizon under a seen channel condition (CDL-A, $100$ ns delay spread, $5$ m/s velocity) at $E_b/N_0=-4\,\mathrm{dB}$.}
            \label{fig: rate_seen}
        \end{minipage}
        \vspace{-4mm}
    \end{figure*}

    \section{Simulation Results}\label{Sec: simulation}
    \textcolor{blue}{In this section, we evaluate the performance of the proposed UV3CP framework through Monte Carlo simulations in terms of frame error rate (FER), NMSE, and achievable rate. We additionally report 90\% confidence intervals for the principal NMSE and achievable-rate comparisons. NMSE is used as the primary predictor-level metric because it directly corresponds to the training objective in \eqref{eq: nmse_loss} and measures future-channel reconstruction accuracy without specifying downstream transmission and receiver processing. FER evaluates the impact of prediction errors on link reliability, while achievable rate quantifies their impact on spectral efficiency when the DL precoder is constructed from predicted future CSI.}
    
    \subsection{Simulation Setup}
    In the simulations, we consider a MIMO-OFDM system operating at a carrier frequency of $3.5$ GHz with a subcarrier spacing of $15$ kHz. The system employs $128$ subcarriers. Each slot consists of $14$ OFDM symbols, and each resource block contains $12$ consecutive subcarriers. We use $4$-QAM for symbol modulation and a low-density parity-check code with code rate $1/2$ for channel coding. \textcolor{blue}{A frame is counted as erroneous if at least one decoded information bit differs from the corresponding transmitted information bit.}
    We use two demodulation reference signal (DM-RS) symbols per slot with predefined pilot patterns compliant with the considered NR resource-grid configuration \cite{5g_standards}. For channel estimation and detection, we apply LS channel estimation, linear interpolation for full CFR reconstruction, and LMMSE equalization.    The channel realizations follow the clustered delay line (CDL) models in \cite{cdl}. Channel data are generated using NVIDIA Sionna \cite{sionna}, which provides realistic propagation environments and channel responses based on physically grounded wireless channel models. The BS is equipped with $N_r \in \{2, 8\}$ antennas, and the UE has $N_t=2$ antennas. We employ a uniform planar array (UPA) with  antenna spacing of $\lambda/2$, where $\lambda$ denotes the wavelength.
    
    We consider a prediction setting in which the first slot is used as the UL observation slot, and the subsequent DL slots are predicted. Since each slot contains $14$ OFDM symbols, the past observation length is set to $P=14$. The prediction horizon $L$ is determined by the number of predicted DL slots, for example, predicting $4$ DL slots gives $L=56$ OFDM symbols. 
    {\color{blue}For the patch-related parameters, the temporal, frequency, and spatial candidate sets are
    $\mathcal{P}_t=\{2,7,14\}$, $\mathcal{P}_f=\{2,4, 8,16\}$, and $\mathcal{P}_s=\{2\}$, respectively.
    The multi-patch training set is constructed as the Cartesian product $\mathcal{P}_{\rm patch} =\mathcal{P}_f\times\mathcal{P}_t\times\mathcal{P}_s$, and the reference patch size is $\mathbf{p}_0=(8,7,2)$.
    For the temporal and frequency-domain patch-selection rules, we set \(L_{t,\max}=L\) and $M_{f,\max}=\max\mathcal{P}_f$.}

    \subsection{Baselines}
    {\color{blue}For performance evaluation, we consider the following baselines.
    \textbf{UV3CP (Proposed)} is the proposed universal ViT framework with statistics-aware patch selection and embedding resizing. It consists of 4 encoder layers with 4 attention heads, an embedding dimension of 128, and an MLP dimension of 256, with $\eta_t=\eta_f=0.1$.
    \textbf{No prediction} directly reuses the most recently estimated channel over the future prediction interval.
    \textbf{LMMSE} is a model-based linear predictor applied independently to each BS-UE subcarrier channel coefficient using the most recent $P=14$ LS channel estimates and Jakes-based temporal correlation.
    \textbf{Kalman filter} employs an AR(1) approximation of the Jakes-based temporal correlation, with process and observation noise statistics determined from the temporal correlation and LS estimation-error variance, respectively.
    \textbf{LSTM} \cite{lstm_cp} uses 2 recurrent layers with a hidden dimension of 512.
    \textbf{Transformer} \cite{transformer_cp} uses a model dimension of 400, 4 attention heads, 5 encoder layers, 2 decoder layers, and an FFN dimension of 882.
    \textbf{LLM4CP} \cite{llm4cp} uses a frozen DistilGPT-2 backbone with an adapter MLP of dimension 2048.
    \textbf{ChannelMamba} \cite{channelmamba} uses a model dimension of 512, 5 encoder layers, 2 decoder layers, an MLP dimension of 512, and a convolution kernel size of 3.
    Finally, \textbf{ViT $(p_f,p_t,p_s)$} \cite{vit} is a fixed-patch ViT with the same backbone configuration as UV3CP but a predetermined 3D patch size $(p_f,p_t,p_s)$, and is used to evaluate sensitivity to patch configuration and the benefit of universal patch adaptation.}

    To evaluate the generalization capability of the neural predictors, we use a mixed-scenario training set combining CDL-A, CDL-B, and CDL-E channels with delay spreads $\{100,300\}$ ns, user velocities $\{5,15,30\}$ m/s, and SNRs from $-4$ to $6$ dB. We generate $18{,}000$ channel samples, using $15{,}000$ for training and $3{,}000$ for validation. All models are trained or fine-tuned on the same input-output channel pairs using the NMSE loss in \eqref{eq: nmse_loss}. AdamW \cite{adamw} is used with a batch size of $32$, a learning rate of $10^{-4}$, and $500$ epochs. Unless otherwise stated, the statistics used for patch selection and embedding resizing in the simulations are model-consistent second-order statistics constructed from the corresponding CDL parameters.

    \subsection{Prediction Accuracy}
    
    
    Fig. \ref{fig: NMSE_seen} evaluates the prediction accuracy under a seen channel condition to assess the in-distribution performance. Fig. \ref{fig: NMSE_seen} shows the NMSE performance under CDL-A with $100$ ns delay spread and $5$ m/s user velocity. As the number of prediction slots increases, the NMSE of all methods increases because the prediction horizon becomes longer and the channel aging effect becomes more severe. Nevertheless, the proposed method achieves the lowest NMSE over the entire slot range, demonstrating its effectiveness even under mixed-scenario training. Compared with the neural baselines, UV3CP maintains a clear performance advantage, especially for longer prediction horizons, indicating that it more effectively captures the learned in-distribution channel dynamics.
    
    
    \textcolor{blue}{Fig. \ref{fig: NMSE_unseen} evaluates the prediction accuracy under unseen channel profiles to assess out-of-distribution generalization. Fig. \ref{fig: NMSE_unseen} shows the NMSE performance under CDL-D with 200 ns delay spread and 10 m/s user velocity, which is excluded from the mixed training distribution. The proposed UV3CP yields the lowest NMSE under this unseen channel condition, indicating that the learned representation and statistics-aware patch adaptation remain effective under channel-profile mismatch. This evaluation demonstrates robustness to the considered unseen channel profile, but does not imply unrestricted generalization to arbitrary channel families or antenna configurations.}
    
    {\color{blue} Fig. \ref{fig: rate_seen} evaluates the achievable rate when the BS constructs the DL precoder from the predicted future channel $\tilde{\mathbf H}_{\rm DL}[n,k]$. 
    The achievable rate is computed as
    \begin{align}
        &R[n,k] = \nonumber \\
        &\log_2\det\!\left(\mathbf I_{N_t}\!+\!\frac{\rho}{N_s\sigma_v^2}\mathbf H_{\rm DL}[n,k]\mathbf F[n,k]\mathbf F^H[n,k]\mathbf H_{\rm DL}^H[n,k]\right),
    \end{align}
    where $\rho$ is the total transmit power at the BS, $\mathbf H_{\rm DL}[n,k]$ is a true DL channel, and $\mathbf F[n,k]$ is the precoding matrix constructed from the predicted DL channel such that $\operatorname{tr}(\mathbf F[n,k]\mathbf F^H[n,k])=N_s$. Prediction errors therefore lead to precoder mismatch with respect to the true future channel, directly reducing the achievable rate. Fig. \ref{fig: rate_seen} shows that the proposed UV3CP achieves a higher achievable rate than the considered model-based and neural channel-prediction baselines over the prediction horizon. As the prediction horizon increases, channel aging increases the mismatch between the predicted and actual future channels and reduces the achievable rate. Nevertheless, UV3CP consistently maintains a higher rate, demonstrating that its improved prediction accuracy also translates into a link-level performance gain.}

    \subsection{Adaptation and Robustness}
    
    \begin{figure}[t]
        \vspace{-2mm}
        \centering
        \includegraphics[width=8cm, trim=12 0 0 0, clip]{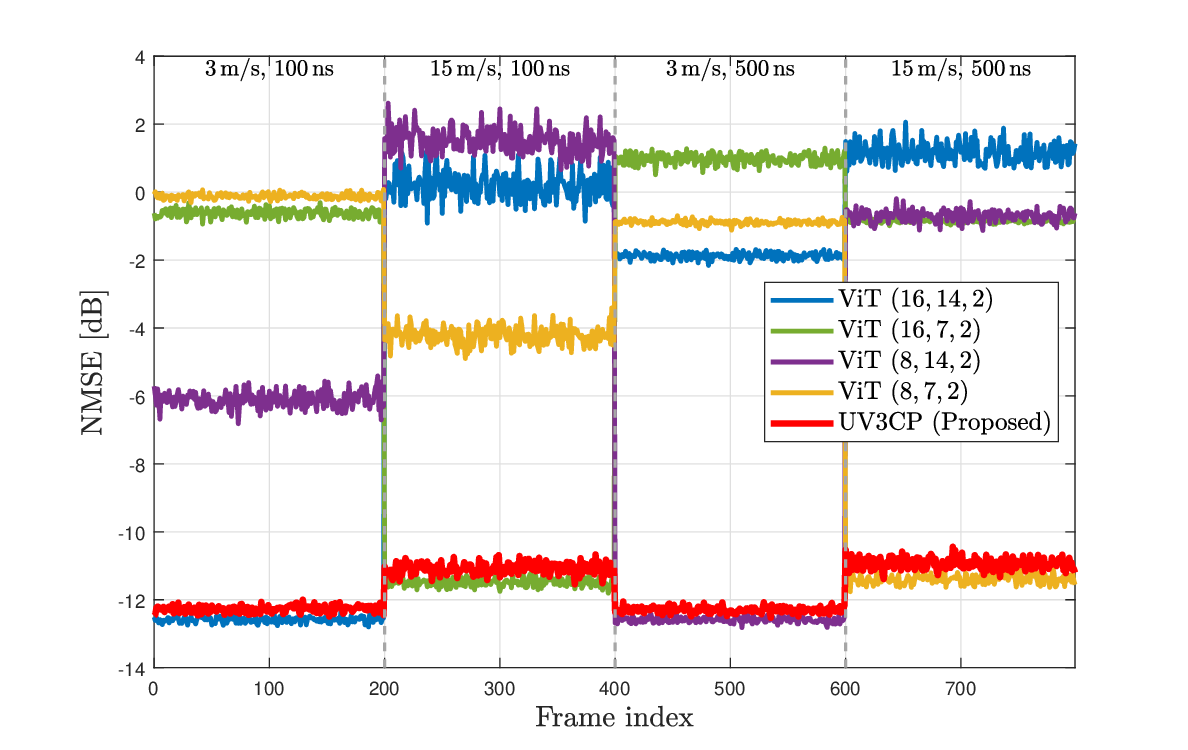}
        \caption{NMSE versus frame index under time-varying channel conditions at $E_b/N_0=2\,\mathrm{dB}$. In each segment, the proposed patch-size selection algorithm determines the patch configuration according to the current channel statistics.}
        \label{fig: NMSE_adaptation}
    \end{figure}
    
    \textcolor{blue}{Fig. \ref{fig: NMSE_adaptation} evaluates the robustness of the proposed UV3CP under time-varying channel conditions. Each interval corresponds to a different velocity and delay-spread setting, and the proposed patch-selection algorithm determines the patch configuration according to the current channel statistics. The fixed-patch ViT models exhibit noticeable performance fluctuations across channel transitions, showing that a patch size that is suitable in one scenario can become suboptimal in another. By contrast, the proposed UV3CP maintains a more stable NMSE profile across the entire sequence. This demonstrates that our statistics-aware patch selection technique effectively tracks scenario-dependent changes in the preferred patch configuration and achieves robust channel prediction in dynamic deployment environments. We further evaluate the practical overhead of this adaptation by measuring the wall-clock reconfiguration latency, including channel-statistics construction, patch-size selection, and patch-embedding resizing. The measured reconfiguration latency is 9.148--32.371~ms. Since reconfiguration is performed once per 200-frame interval, this overhead is amortized to only 0.0457--0.1619~ms per frame, corresponding to 1.60--3.28\% of the fixed-patch inference latency.}

    
    \begin{figure}[t]
        \vspace{-4mm}
        \centering
        \includegraphics[width=8cm]{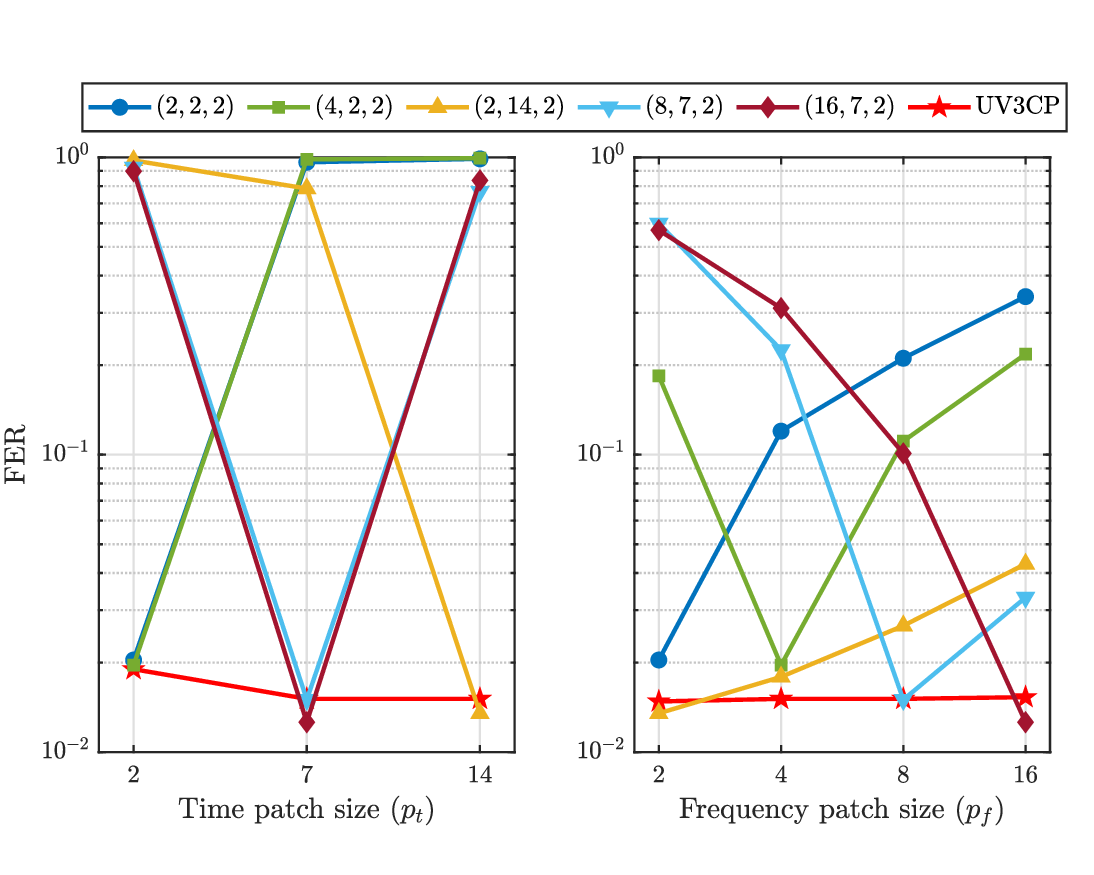}
        \caption{FER versus temporal and frequency patch sizes for fixed-patch ViT models and the proposed UV3CP at $E_b/N_0=10\,\mathrm{dB}$.}
        \label{fig:FER_mismatch}
        \vspace{-5mm}
    \end{figure}
    
    {\color{blue} Fig. \ref{fig:FER_mismatch} compares the FERs of fixed-patch ViT models and the proposed UV3CP under axis-wise changes in the test patch size. The left and right plots vary the temporal patch size $p_t$ and frequency patch size $p_f$, respectively. Each fixed-patch ViT is trained with the patch size indicated in the legend. When evaluated with a different test patch size, its learned patch-embedding kernel is resized to the target patch dimension for dimensional compatibility, while the remaining model parameters are kept fixed. In contrast, UV3CP employs the statistics-aware patch adaptation strategy. The fixed-patch ViTs exhibit noticeable FER degradation under patch-size mismatch, whereas UV3CP maintains consistently low FER across both temporal and frequency patch-size variations. These results demonstrate the robustness of the proposed embedding resizing and patch adaptation strategy to patch-size mismatch.}
    
    \subsection{Ablation Study and Complexity}
    
    
    \begin{table}[t]
        \centering
        \caption{NMSE [\textnormal{dB}] comparison of the proposed statistical resizing and baseline resizing methods under different $E_b/N_0$ values.}
        \label{tab:resize_comparison}
        \begin{tabular}{c|ccccc}
        \toprule
        $E_b/N_0$ [dB] & Fixed & Lin. & Lin.-pinv & Nearest & Statistical \\
        \midrule
        -10 &  0.073 & -0.477 & -0.477 & -0.438 & \textbf{-0.487} \\
         -6 & -2.950 & -3.969 & -3.925 & -3.947 & \textbf{-4.012} \\
         -2 & -5.346 & -7.033 & -6.946 & -6.925 & \textbf{-7.077} \\
          2 & -7.122 & -9.508 & -9.318 & -9.281 & \textbf{-9.547} \\
          6 & -8.239 & -11.024 & -10.862 & -10.706 & \textbf{-11.249} \\
         10 & -8.827 & -11.938 & -11.675 & -11.487 & \textbf{-12.218} \\
        \bottomrule
        \end{tabular}
    \end{table}
    
    {\color{blue}Table \ref{tab:resize_comparison} compares the NMSEs of different resizing methods under various $E_b/N_0$ values. In this comparison, \emph{Fixed} directly reuses the reference patch embedding without resizing, whereas \emph{Lin.}, \emph{Lin.-pinv}, and \emph{Nearest} are non-statistical baselines based on interpolation or pseudoinverse correction. In particular, \emph{Lin.-pinv} corresponds to the pseudoinverse-based resize strategy of FlexiViT, implemented using the same shared ViT backbone, multi-patch training protocol, training data, and optimization settings as the proposed method. Thus, the comparison with \emph{Lin.-pinv} isolates the effect of replacing the pseudoinverse-based resizing rule with the proposed statistics-aware resizing. The proposed statistical resizing achieves the lowest NMSE over the entire tested $E_b/N_0$ range, consistently improving upon \emph{Lin.-pinv} and the other non-statistical alternatives. This result indicates that incorporating second-order channel statistics into the resizing procedure provides more effective patch adaptation than purely geometric resizing rules.}
    
    \begin{table}[t]
        \centering
        \setlength{\tabcolsep}{5pt}
        \caption{NMSE [\textnormal{dB}] comparison of attention configurations under different $E_b/N_0$ values.}
        \label{tab:attention_comparison}
        \begin{tabular}{c|cccc}
        \toprule
        $E_b/N_0$ [dB] & F & T & FT & FTA \\
        \midrule
        -10 & -0.434 & -0.209 & -0.521 & \textbf{-0.555} \\
         -6 & -3.925 & -3.925 & -4.023 & \textbf{-4.056} \\
         -2 & -7.100 & -7.077 & -7.190 & \textbf{-7.212} \\
          2 & -9.666 & -9.547 & -9.747 & \textbf{-9.788} \\
          6 & -11.427 & -11.192 & -11.487 & \textbf{-11.549} \\
         10 & -12.441 & -12.076 & -12.441 & \textbf{-12.596} \\
        \bottomrule
        \end{tabular}
        \vspace{-3mm}
    \end{table}
    
    Table \ref{tab:attention_comparison} compares the NMSEs of different combinations of the attention domains under various $E_b/N_0$ values. For a fair comparison, all configurations use the same statistics-aware patch-size selection and embedding resizing procedures; only the attention domains are varied. Here, \emph{F}, \emph{T}, and \emph{A} denote frequency-, temporal-, and spatial-domain attention, respectively, while \emph{FT} and \emph{FTA} denote their corresponding combinations. The FTA configuration consistently achieves the lowest NMSE over the entire tested $E_b/N_0$ range. In particular, the consistent improvement of FTA over FT indicates that spatial-domain attention provides additional gains beyond joint frequency-temporal modeling. This confirms the benefit of jointly exploiting the frequency, temporal, and spatial correlations of the channel.

    \begin{table}[t]
        \centering
        \caption{Model size, storage, and forward-pass complexity for supporting patch configurations.}
        \label{tab:model_complexity}
        \footnotesize
        \setlength{\tabcolsep}{3pt}
        \begin{tabular}{c|cc|cc|c}
            \hline
            Prediction & \multicolumn{2}{c|}{Parameters [million]} & \multicolumn{2}{c|}{Storage [MiB]} & Forwards \\
            length $(L)$ & UV3CP & 12 fixed ViTs & UV3CP & 12 fixed ViTs & GFLOPs \\
            \hline
            $14$ & 31.175  & 374.102  & 119.240 & 1430.879 & 0.782 \\
            $56$ & 119.944 & 1439.324 & 457.865 & 5494.379 & 2.158 \\
            $98$ & 208.712 & 2504.546 & 796.490 & 9557.879 & 3.535 \\
            \hline
        \end{tabular}
        \vspace{-5mm}
    \end{table}
    {\color{blue}
    Table \ref{tab:model_complexity} evaluates the model size, storage, and forward-pass complexity of different channel predictors.
    Table \ref{tab:model_complexity} shows that UV3CP reduces the aggregate parameter count and checkpoint-storage requirement by approximately $91.7\%$ compared with maintaining 12 separately trained fixed-patch ViTs. This reduction follows from replacing a bank of patch-specific models with a single shared predictor. The resized patch-embedding kernels do not require separate model checkpoints, and even caching the kernels for all candidate patch configurations introduces only a negligible additional memory cost. For a given patch configuration, UV3CP and the corresponding fixed-patch ViT use the same token grid and prediction backbone. Therefore, their forward-pass complexity is identical, and the last column of Table \ref{tab:model_complexity} reports this common per-inference cost. The forward complexity increases with the prediction horizon because the output reconstruction head grows with the number of future channel samples to be predicted. Thus, the main complexity advantage of UV3CP lies not in reducing the per-inference backbone cost, but in supporting multiple patch configurations with a single trained and stored model.}
    
    \section{Conclusion}\label{Sec: conclusion}
    In this paper, we proposed a universal ViT-based 3D channel prediction framework, called {\em UV3CP}, for MIMO-OFDM channel prediction in TDD systems. The proposed UV3CP addresses the practical limitation of patch-size sensitivity in ViT-based predictors by combining a statistics-aware patch-size selection with a universal ViT enabled by statistics-aware patch embedding resizing. Simulation results demonstrated that the proposed UV3CP achieves improved FER, NMSE and rate performance and provides stronger robustness to patch-size mismatch and scenario variation than fixed-patch ViT models. Overall, the results highlight the importance of both scenario-adaptive patch design and universal patch-size adaptation for reliable ViT-based channel prediction in dynamic wireless channels.

\end{document}